\documentstyle[aps,eqsecnum,epsf,floats]{revtex}
\begin{document}

\begin{center}
{\Large \bf Double  Distributions and Evolution Equations  }
\end{center}
\begin{center}
{A.V. RADYUSHKIN\footnotemark
}  \\
{\em Physics Department, Old Dominion University,}
\\{\em Norfolk, VA 23529, USA}
 \\ {\em and} \\
{\em Jefferson Lab,} \\
 {\em Newport News,VA 23606, USA}
\end{center}
\vspace{2cm}

\footnotetext{Also Laboratory of Theoretical  Physics, JINR, Dubna, Russian Federation}

\begin{abstract}

Applications of perturbative QCD to
deeply virtual Compton scattering
and hard exclusive meson electroproduction 
processes require a generalization of 
usual parton distributions for the case when
long-distance information
is accumulated in  nonforward matrix
elements $\langle p' \,  |     \,   
{\cal O}(0,z)  \,  |     \, p \rangle $ 
of quark and gluon light-cone  operators.
In our previous papers we used   
two types of nonperturbative functions 
parametrizing such matrix elements:
double distributions $F(x,y;t)$
and nonforward distribution functions
${\cal F}_{\zeta}(X;t)$.    
Here we discuss in   more detail 
the  double distributions (DDs)
 and evolution equations which they satisfy.
We propose simple models for $F(x,y;t=0)$ DDs 
with correct  
spectral and symmetry properties
which also satisfy the 
 reduction relations connecting them to the 
usual  parton densities $f(x)$.
In this way, we obtain self-consistent models for 
the $\zeta$-dependence of nonforward distributions.
We show that, for small $\zeta$, one can easily 
obtain nonforward distributions (in the $X > \zeta$ region)
from the  parton densities:
${\cal F}_{\zeta}(X;t=0) \approx f(X - \zeta/2)$.

\vspace{5mm}

PACS number(s): 12.38.Bx, 13.60.Fz, 13.60.Le

\end{abstract}

\newpage

\section{Introduction}

Applications of perturbative QCD to
deeply virtual Compton scattering 
and hard exclusive electroproduction 
processes \cite{ji,compton,ji2,gluon,npd,bfgms,cfs} 
require a generalization of 
usual parton distributions for the case when
long-distance information
is accumulated in  nonforward matrix
elements $\langle p-r \,  |     \,   
{\cal O}(0,z)  \,  |     \, p \rangle \,  |     \, _{z^2=0}$ 
of quark and gluon light-cone  operators.
As argued in Refs.  \cite{compton,gluon,npd},
 such matrix elements can be parametrized by 
two basic  types of nonperturbative functions.
With $z$ taken in the lightcone ``minus'' direction, 
the   double distributions (DDs) $F(x,y;t)$ 
specify the light-cone ``plus'' fractions $xp^+$ and $yr^+$ 
of the initial hadron momentum $p$ and the momentum transfer $r$ 
carried by the initial parton. 
Though $z$ is an integration variable,
only one direction on the lightcone (specified by external
momenta) is important for the lightcone-dominated processes.
In other words, only 
the lightcone plus  direction of the hadron 
$p$ and $r$ momenta are essential 
for such processes.  
By definition,  the DDs $F(x,y;t)$ do not depend 
on the $r^+/p^+$ ratio. 
On the other hand, treating the proportionality  coefficient as an
independent parameter:
$r^+  \equiv \zeta p^+$,  one can   introduce 
an alternative description in terms
of the   nonforward parton distributions 
${\cal F}_{\zeta}(X;t)$ 
with $X=x+y \zeta$ being the total 
fraction of the initial hadron momentum 
taken  by the initial  parton.
The shape of the functions ${\cal F}_{\zeta}(X;t)$ explicitly
depends on the parameter $\zeta$ characterizing the skewedness
of the relevant nonforward matrix element.
This parametrization of  nonforward matrix 
elements  by ${\cal F}_{\zeta}(X;t)$
is similar to that proposed originally by 
X. Ji \cite{ji,ji2} who introduced 
 off-forward parton distributions (OFPDs) $H(x,\xi;t)$. 
The latter are close to functions considered 
earlier in Ref.  \cite{drm}. 
The functions   
 $H(x,\xi;t)$ have a simple   relation  to  
nonforward distributions (NFPDs)  
${\cal F}_{\zeta}(X;t)$, while  the 
 non-diagonal distributions  $F(x_1,x_2)$ 
discussed by  Collins, Frankfurt 
and Strikman \cite{cfs}  essentially coincide with 
${\cal F}_{\zeta = x_1 - x_2}(x_1;t=0)$
(see Ref. \cite{npd} for details).
The  basic distinction between our 
approach   and those  of Refs. \cite{ji,ji2,cfs}
is that we treat the  
  double distributions $F(x,y;t)$ 
as the primary objects of the QCD analysis
producing the nonforward distributions ${\cal F}_{\zeta}(X;t)$ 
(and other types of distributions) after
an appropriate  integration. 

The formalism  of double distributions 
provides  a rather effective tool 
for   studying   
some general (e.g., spectral)  properties of NFPDs and  
it allows  to find analytic  solutions 
of    evolution equations \cite{compton,gluon,npd}.
Incorporating symmetries of DDs \cite{lech}
imposes  rather strong restrictions on 
  realistic models of NFPDs. 
A possible strategy for a self-consistent model building is to use 
nonperturbative or phenomenological
approaches (MIT bag,   
quark models, QCD sum rule ideas, etc.)
to construct double distributions
at low normalization point and then evolve them to higher
$Q^2$ values.  

The evolution equation  
for the nonsinglet quark double distribution  
was derived in Ref.\cite{compton}, where its analytic  solution
was also given. Evolution of the gluon distribution
in pure gluodynamics was discussed in Refs.\cite{gluon,npd}.
In this paper, we present also a  full set of evolution
equations for the flavor-singlet case 
and derive a  solution following 
the method of Refs.\cite{compton,gluon,npd}. 
An  independent study of singlet evolution 
based on our approach was performed  
in Ref.\cite{bgrdouble}. 
Evolution equations for various versions 
of nonforward distributions can be found 
in \cite{ji2,npd,ffgs,bgr,gevol}.
 A convenient way to obtain  the relevant evolution  kernels
 is to use the  universal
light-ray evolution kernels  \cite{bgr,gevol,gdr,bb}. 
The evolution 
of  nonforward distributions 
was studied numerically in refs.\cite{ffgs,bel,lech,maryskin,gokmar}. 

In the present paper, we incorporate  
the spectral and symmetry 
properties of double distributions
to construct some simple models for DDs. 
Using  the relations 
between DDs and 
NFPDs/OFPDs, we derive models for the latter
and show
that using  the formalism of 
double distributions we can  easily  explain 
characteristic  qualitative and quantitative 
features of the evolution
of nonforward  distributions observed in 
Refs.\cite{ffgs,maryskin}.

\section{Basic definitions}

The  kinematics of the 
amplitudes of the DVCS process $\gamma^*(q) N(p) \to \gamma(q')  N(p')$ 
and hard electroproduction 
$\gamma^* (q) N(p) \to M (q')  N(p')$ can be 
 specified  
by the initial nucleon momentum $p$,
the   momentum  transfer $r=p-p'$ and  the 
momentum $q'$  of the final photon or meson. 
To get a  Bjorken-type  scaling limit, 
one should also keep the invariant momentum transfer 
$t \equiv r^2$  
 small   compared to
the  virtuality $-Q^2 \equiv (q-r)^2$ of the initial 
photon and the energy invariant $p \cdot q \equiv m_p \nu$. 
The  
essential features of the hard 
electroproduction processes (DVCS included)
 can be most easily demonstrated if we 
 set $q'^2=0$,  $p^2=0$,  $r^2 = 0$  and 
use $p$, $q'$  as the  basic light-cone  (Sudakov) 4-vectors.
It is easy to see that  the  requirement
$p'^2 \equiv (p+r)^2=p^2$  reduces in this limit to the condition 
$p\cdot r = 0$  which  can be satisfied only
if the  two lightlike momenta
$p$ and $r$ are proportional to each other: $r= \zeta  p$,
where $\zeta$ coincides with 
the Bjorken variable $\zeta = x_{Bj} \equiv Q^2/2(p \cdot q)$. The latter 
 satisfies the constraint  $0 \leq x_{Bj}  \leq 1$.
For small but finite $t$ and $m_p$, the momentum transfer
$r$ still must have a non-zero plus component
$r^+= \zeta p^+$. It also may have  a transverse component $r_{\perp}$.

In the pQCD factorization treatment of hard
electroproduction processes, the 
nonperturbative information is accumulated in the
nonforward matrix element
$\langle p-r \,  |     \,  \varphi (0) \varphi (z) \,  |     \,  p \rangle $
(we use here $\varphi$ as a generic notation for 
quark ($\psi$) or gluonic ($G$) fields). 
It  depends on the relative coordinate $z$
through three invariant variables $(pz), (rz)$ and $z^2$.
In the forward case, when $r=0$, 
one gets the usual  parton distributions 
by Fourier transforming the light-cone 
projected (i.e., $z^2=0$) matrix element 
with respect to $(pz)$.
In the  nonforward case, we can  
try to start with the general  
Fourier  representation 
\begin{equation}
\langle p-r \,  |     \,  \varphi (0) \varphi (z) \,  |     \,  p \rangle 
\equiv {\cal M} ((pz),(rz),z^2; t,m_p^2) 
= \int_{- \infty}^{\infty}dx \, \int_{- \infty}^{ \infty} dy \, 
\int_{- \infty}^{\infty} e^{-ix(pz) -iy(rz)} 
\rho(x,y,\nu;t) \,e^{-iz^2 \nu} d \nu 
\label{21} \end{equation}
with respect to all three $z$-dependent invariants.
The Fourier transform $\rho (x,y,\nu;t)$
can 
be called a {\it triple} distribution. 
Note that the generous $({- \infty}, {\infty})$ limits 
for all three variables $x,y, \nu$  serve for a most general 
function of $(zp), (zr) $ and $z^2$. 
However, incorporating information that
the Fourier transformation is written for
a  function ${\cal M}$ given by 
 Feynman integrals having specific causality properties,  
one arrives at more narrow  limits:  $\nu$ runs from 0 to
$\infty$, $x$ is  between $-1$  and 1 while  $y$ is 
 between $0$  and 1
 (this was  proven in \cite{npd}
for any Feynman 
diagram using the approach of Ref.
\cite{spectral}). 
To interpret the $x$-variable  as the fraction
of the initial momentum $p$ carried by the
relevant parton, it makes sense to separate
integration over  positive and 
negative $x$ components and redefine $x \to -x $   
and $y \to 1-y$ for the negative $x$ component.
After that, the $x$-variable is 
always positive and   
 $x$ and $y$ are further constrained
 by inequality 
$0 \leq x+y \leq 1$ \cite{compton,npd}. 
 These spectral conditions
can be summarized  by  the following representation
  \begin{eqnarray}
&& \lefteqn{  \langle p-r \,  |     \,  \varphi (0) \varphi (z) \,  |     \,  p \rangle 
=} \label{22}
\\ && \hspace{1.5cm}   \int_{0}^{\infty} d \nu \, e^{-i\nu 
(z^2 -i \epsilon)} 
 \int_{0}^{1} \int_{0}^{1}
\biggl [ e^{-ix(pz) -iy(rz)} 
\Phi(x,y,\nu;t) +  e^{ix(pz) - 
i(1-y)(rz)}\bar \Phi(x,y,\nu;t) \biggl ]
 \theta(x+y \leq 1)  \,dx \, dy  \, , \nonumber 
\end{eqnarray}
in which   
$\Phi(x,y,\nu;t)$ and $\bar \Phi(x,y,\nu;t)$ 
result from  positive-$x$ and negative-$x$ components 
of $\rho (x,y,\nu;t)$
respectively.  In particular, for quark 
operators, 
$\Phi(x,y ,\nu;t) $ can be interpreted as the 
quark distribution while $\bar \Phi(x,y,\nu ;t) $
 as the antiquark one (a more detailed discussion is given
 in the next section). 
 Similarly, $y$ and $(1-y)$
can be interpreted as the fractions in which 
 the  momentum transfer $r$  is shared among the
two fields of the composite operator $\varphi(0) \varphi (z)$.
Finally, the  $\nu$ variable characterizes
the virtuality of these fields.
For a light-cone dominated process, 
the leading term is given by the 
 $z^2 \to 0$ limit of the 
nonforward matrix element, i.e. by 
zeroth moment of $ \Phi(x,y,\nu;t )$ with respect
to $\nu$
\begin{equation}
F(x,y;t ) = \int_0^{\infty}  \Phi(x,y,\nu;t)\, d \nu \,  ,
\label{23} \end{equation}
where $ F(x,y;t )$ is the
{\it double distribution}.

For a lightlike interval $z^2=0$, one can treat
$z$ as having only light-cone ``minus'' component,
and then  the scalar products $(pz), (rz)$ project out the ``plus''
components of general (non-lightlike) momenta $p$ and $r$. 
This  allows to give a  parton interpretation 
of  $ F(x,y;t )$ as a probability amplitude for the
active parton to carry fractions $xp^+$
and $yr^+$ of the plus components of the external 
momenta $r$ and $p$.
 Though  the momenta
$p^+$ and $r^+$ can be treated  as
proportional to each other $r^+ = \zeta  p^+$,
 $p^+$ and $r^+$  specify 
the ``+''-momentum flow in two different   channels.
For $r^+=0$, the net ``+''-momentum flows 
only in the $s$-channel 
and the total ``+''-momentum entering into 
the  composite operator vertex is zero.
In this case, 
the matrix element is analogous to a
 distribution function.
The  partons  entering the composite vertex 
then carry  the   fractions $x_i p^+$ 
of the initial proton momentum ($-1 < x_i <1$). 
When  $x_i $ is  negative, we  interpret the parton 
as belonging to the final state to  secure that  
the integral always  runs over  the segment  $0\leq x\leq 1$.
In this parton picture, the spectators take the 
remaining momentum  $(1-x)p^+$.
On the other hand, if  the 
total ``+''-momentum flowing through the 
composite vertex is $r^+$,
the matrix element has the structure
of a distribution amplitude in which 
the momentum  $r^+$  
splits into the fractions $yr^+$  and 
$(1-y)r^+ \equiv \bar y r^+$ carried by the 
two fields  that appear in  the vertex. 
In a combined situation, when both $p^+$ and $r^+$
are nonzero,  the initial parton  takes 
$xp^+ +y r^+$, while the final one  carries the momentum 
$xp^+ - \bar y r^+$. 
For $r=0$, we get the forward matrix element
which is parametrized by the usual parton distributions
$f(x)$.
This gives  reduction relations 
 \cite{compton,gluon,npd}
connecting  double distributions with the usual ones
(see Eqs.(\ref{34}),(\ref{35}) below).

\section{Quark and gluon distributions}

For quark operators, 
the double distributions are defined by the
 following representation \cite{compton}:
\begin{eqnarray} 
&& \langle p',s'\, \,  |     \,  \, \bar \psi_a(0) \hat z 
E(0,z;A)  \psi_a(z) \, \,  |     \,  \, p,s \rangle \,  |     \, _{z^2=0} 
\label{31}  \\ &&  =  \bar u(p',s')  \hat z u(p,s)  
\int_0^1    \int_0^1  \,  
 \left ( e^{-ix(pz)-iy(r z)}F_a(x,y;t) 
  -   e^{ix(pz)-i\bar y(r z)}F_{\bar a}(x,y;t)
\right )\theta( x+y \leq 1) \, dx \, dy 
 \nonumber \\ &&
+ \frac1{4M} \, \bar u(p',s')  (\hat z \hat r - \hat r \hat z)
u(p,s)  \int_0^1   \int_0^1  \,
  \left ( e^{-ix(pz)-iy(r z)}K_a(x,y;t) 
  -   e^{ix(pz)-i\bar y(r z)}K_{\bar a}(x,y;t)
\right ) \theta( x+y \leq 1) \,dx\, dy  \   \,
\nonumber 
 \end{eqnarray} 
for parton helicity-averaged ones and by 
\begin{eqnarray} 
&& \langle p',s'\, \,  |     \,  \, \bar \psi_a(0)  \hat z \gamma_5
E(0,z;A)  \psi_a(z) \, \,  |     \,  \, p,s \rangle \,  |     \, _{z^2=0}  
 \label{32}  \\ && =  \bar u(p',s')  \hat z  \gamma_5 u(p,s)  
\int_0^1   \int_0^1  \,  
 \left ( e^{-ix(pz)-iy(r z)}G_a(x,y;t) 
+  e^{ix(pz)-i\bar y(r z)}
G_{\bar a}(x,y;t) \right )  \theta( x+y \leq 1) \, dx \, dy 
 \nonumber \\ &&
+ \frac{(rz)}{2M} \bar u(p',s')  \gamma_5 u(p,s)  
\int_0^1   \int_0^1  \, 
  \left ( e^{-ix(pz)-iy(r z)}P_a(x,y;t) 
  +   e^{ix(pz)-i\bar y(r z)}P_{\bar a}(x,y;t)
\right ) \theta( x+y \leq 1) \, dx \, dy 
\nonumber 
 \end{eqnarray}
in the parton helicity-sensitive case. 
Here and in what follows we adhere to  the ``bar'' convention
$\bar y = 1-y, \bar x = 1-x$, $etc.,$ 
 for  momentum fractions and use the ``Russian hat'' notation
$\gamma_{\alpha} z^{\alpha} \equiv \hat z$.
As usual, 
 $\bar u(p',s'), u(p,s)$ are the Dirac spinors for the nucleon.
In this definition, we explicitly separate 
quark and antiquark components
of  the double distribution.
 Note that such a  separation is unambiguous:
in the Fourier representation, it is completely determined by the sign 
of the $x$-term in the exponential.

To clarify the physical meaning of separating 
the DDs into two components, 
it is  instructive to consider the forward limit $r=0$ 
in which the matrix element is parametrized by 
usual parton densities, e.g., in the helicity averaged case
\begin{equation} 
\langle p\, \,  |     \,  \, \bar \psi_a(0) \hat z 
E(0,z;A)  \psi_a(z) \, \,  |     \,  \, p \rangle \,  |     \, _{z^2=0} 
 =  \bar u(p)  \hat z u(p)  
   \int_0^1  \,  
 \left ( e^{-ix(pz)}f_a(x) 
  -   e^{ix(pz)}f_{\bar a}(x)
\right ) \, dx \, . 
\label{33} \end{equation} 
The exponential factors accompanying
the quark and antiquark distributions 
reflect the  fact that the field $\psi (z) $
appearing in the  operator
$\bar \psi (0)  \ldots \psi (z) $  
consists of  the quark annihilation 
operator (quark with momentum $xp$
comes into  this point) and the antiquark
creation operator (i.e., antiquark with momentum
$xp$ goes out of this point).
To get the relative signs with which  quark 
and antiquark distributions appear in these definitions,
we should take into account 
that antiquark creation and annihilation
operators appear in $\bar \psi (0)  \ldots \psi (z) $ 
in the opposite  order.
Comparing the expression (\ref{33}) 
with the $r =0$ limit of the definitions for DDs,
we obtain   
``reduction formulas'' 
relating   the two components of the double distributions 
to the quark and antiquark parton  densities, respectively:
\begin{equation} 
\int_0^{1-x} \, F_{a}(x,y;t=0)\, dy= 
 f_{a}(x) \hspace{1cm} ; \hspace{1cm} 
\int_0^{1-x} \, F_{\bar a}(x,y;t=0)\, dy= 
 f_{\bar a}(x) \, , \label{eq:redfsym}
 \label{34} \end{equation}
and similarly for the  helicity-sensitive case:
\begin{equation} 
\int_0^{1-x} \, G_{a}(x,y;t=0)\, dy= 
 \Delta f_{a}(x) \hspace{1cm} ; \hspace{1cm} 
\int_0^{1-x} \, G_{\bar a}(x,y;t=0)\, dy= 
\Delta  f_{\bar a}(x)  \,  . 
\label{35} \end{equation}
The reduction formulas 
tell us   that integrating 
the double distribution  $F_{a}(x,y;t=0)$ 
over a vertical line $x=$const in the $(x,y)$-plane,
one gets the quark density $f_{a}(x)$ while 
integrating 
its counterpart    $F_{\bar a}(x,y;t=0)$
gives the  antiquark density $f_{\bar a}(x)$.  
This is an  illustration of   our statement 
that  $F_{a}(x,y)$ and $F_{\bar a}(x,y)$
are independent 
 functions. In particular,
 $F_{a}(x,y)$
 contains the valence component 
(reducing to $f_a^{val}(x)$)  absent in $F_{\bar a}(x,y)$.

 Our definitions (\ref{31}), (\ref{32}) 
 reflect  the results of the $\alpha$-representation
 analysis \cite{npd} that the plus component 
 of the momentum of the  particle (either quark or antiquark) 
 {\it going out}  of the
hadronic blob  can be written as $xp^+ + yr^+$ 
with both $x$ and $y$ positive and $x+y \leq 1$.
 This is in full compliance with 
 the  parton model based expectation that  
 the initial 
 hadron splits into an  active parton and spectators
which  both carry positive fractions of its plus momentum.
To show  the positivity of the plus momentum
component for spectators,
we should explicitly take into account that, in  the 
kinematics of  DVCS and hard 
electroproduction processes, the plus component of the
momentum transfer
$r=p-p'$  is positive $r^+ = \zeta p^+ >0$.
Requiring that the  plus component of the
final hadron momentum is also positive,
we conclude that $0 \leq \zeta  \leq 1$.
Hence, $0 \leq x+y \zeta \leq 1$ (since $0 \leq \zeta  \leq 1$
and $0 \leq x+y \leq 1$), i.e., the plus component of the 
momentum carried by spectators 
is  also  positive.  
On the other hand, the parton  {\it ``going back''}
 has the momentum whose plus component 
$xp^+ - \bar yr^+ = (x - \bar y \zeta)p^+ $
may be either positive or negative, depending on the 
relationship between $x$, $y$ and $\zeta$. 
When $(x - \bar y \zeta) $ is negative,
one may wish to interpret such a  parton 
as an antiparton leaving  the hadron 
together with the initial parton. 
One should remember, however, that the double distributions 
$F(x,y;t)$  ``know nothing'' about the magnitude of 
the skewedness  $\zeta$: 
they are universal functions 
describing flux of $p^+$ and $r^+$ 
independently of what the ratio $r^+/p^+$ might be. 
As we explained above, the quark DDs  are 
 unambiguously divided  into two separate
components $F_a(x,y;t)$ and  $F_{\bar a}(x,y;t)$,
but there is no further subdivision inside them based on 
interrelation between the values of 
$x$ and $y$.

In a similar way, we can introduce  
 double distributions for the gluons
\begin{eqnarray} 
&& \langle p',s'  \,  \,  |     \,  \,   
z_{\mu}  z_{\nu} G_{\mu \alpha}^a (0) E_{ab}(0,z;A) 
G_{ \alpha \nu}^b (z)\, \,  |     \,  \,p,s  \rangle \,  |     \, _{z^2=0}
  \label{36}  \\  &&  
= \bar u(p',s')  \hat z 
 u(p,s) \, (z \cdot p) \int_0^1   \int_0^1  \, 
  \frac1{2} \left( e^{-ix(pz)-iy(r z)} 
+e^{ix(pz)-i\bar y(r z)}\right ) 
\theta( x+y \leq 1) \,  F_g(x,y;t)  
 \, x dx \, dy  + ``K_g"{\rm -term}.
\nonumber
 \end{eqnarray} 

\begin{eqnarray} 
&& \langle p',s'  \,  \,  |     \,  \,   
z_{\mu}  z_{\nu} G_{\mu \alpha}^a (0) E_{ab}(0,z;A) 
\tilde G_{ \alpha \nu}^b (z)\, \,  |     \,  \,p,s  \rangle \,  |     \, _{z^2=0}
  \label{37}  \\  &&  
= \bar u(p',s')  \hat z \gamma_5 
 u(p,s) \, (z \cdot p) \int_0^1   \int_0^1  \, 
  \frac{i}{2} \left( e^{-ix(pz)-iy(r z)} 
-e^{ix(pz)-i\bar y(r z)}\right ) 
\theta( x+y \leq 1) \,  G_g(x,y;t)  
 \,  x dx \, dy  + ``P_g"{\rm -term}.
\nonumber
 \end{eqnarray} 
There are no ``antigluons'',
so the  positive-$x$ and negative-$x$  parts 
are described by the same function.
Note that our definition
of the gluon double distributions here differs   
 from that used in our earlier papers\cite{gluon,npd,gevol}
by an extra factor of $x$ in its right hand side.
This form is more convenient for applications of  the method 
of Refs. \cite{compton,gluon,npd} 
to  solve evolution equations for double distributions
in the singlet case. 
The choice made above  corresponds also 
to the simplest form of the  reduction formulas 
\begin{equation}
\int_0^{1-x} \, F_{g}(x,y;t=0)\, dy= 
 f_{g}(x) \hspace{1cm} ; \hspace{1cm} 
\int_0^{1-x} \, G_{g}(x,y;t=0)\, dy = 
 \Delta f_{g}(x).
\label{38} \end{equation}

Another ambiguity in the definition of the gluon 
double distribution is related to  the overall 
factor $(z \cdot p)$ in the rhs of Eqs.(\ref{36}) and (\ref{37}).
Instead of  it, we could take, e.g., $(z \cdot p')$ or $(z \cdot r)$ 
(such a choice is utterly inconvenient
for taking the forward limit) or $(z \cdot P)$
where $P=(p+p')/2$ is a symmetric 
combination  of the initial and final momenta. 
The latter choice (made in ref.\cite{lech}) is more convenient 
for the studies of symmetry properties of the gluon DDs.
Our  choice made in Eqs. (\ref{36}) and (\ref{37}) simplifies  the
expressions  for off-diagonal
($QG$ and $GQ$)  evolution kernels  
 (see Eq.(\ref{420}) below).

The  
 flavor-singlet 
quark operators
\begin{equation}
{\cal O}_{ Q}(uz,vz) =   \sum_{a=1}^{N_f} 
 \frac{i}{2}
\biggl [ \bar \psi_a(uz) 
\hat z E(uz,vz;A)  \psi_a(vz)
- \bar \psi_a(vz)  \hat z  E(vz,uz;A) \psi_a(uz) \biggr ]   
  \label{39} \end{equation}
and 
\begin{equation}
\Delta {\cal O}_{ Q}(uz,vz) =   \sum_{a=1}^{N_f} 
 \frac{1}{2}
\biggl [ \bar \psi_a(uz) 
\hat z \gamma_5 E(uz,vz;A)  \psi_a(vz)
+ \bar \psi_a(vz)  \hat z \gamma_5 E(vz,uz;A) \psi_a(uz) \biggr ] \,  \label{310} \end{equation}
are expressed in terms of   double  
 distributions $F_Q (x,y;t)$, $G_Q (x,y;t)$, etc. 
specified by  
\begin{eqnarray} 
&& \langle \,  p',s' \,  |     \,  \, {\cal O}_{ Q}(uz,vz)\, 
\,  |     \,  \, p,s \rangle \,  |     \, _{z^2=0} 
= 
 \bar u(p',s')  \hat z  u(p,s)   
\int_0^1   \int_0^1  \, 
  \frac{i}{2} \left( e^{-ix v (pz)-iy  v (r z) +ix
 u(pz)-i\bar y u (r z) } \right. \nonumber \\ && \left. \hspace{2cm} 
-\,  e^{ixv (pz)-i\bar y v(r z) - ix
 u (pz)-i u (r z)  } \right ) 
 \,  F_Q(x,y;t) \, 
 \theta( x+y \leq 1)\,  dx \, dy
  + ``K_Q" {\rm -term}, 
\label{311} \end{eqnarray} 
\begin{eqnarray} 
&& \langle \,  p',s' \,  |     \,  \, \Delta {\cal O}_{ Q}(uz,vz)\, \,  |     \, 
 \, p,s \rangle \,  |     \, _{z^2=0}  = 
 \bar u(p',s')  \hat z \gamma_5 u(p,s)   
\int_0^1   \int_0^1  \, 
  \frac1{2} 
\left( e^{-ix v (pz)-iy  v (r z) +ix
 u(pz)-i\bar y u (r z) } 
 \right. \nonumber \\ && \left.\hspace{2cm} 
+ e^{ixv (pz)-i\bar y v(r z) - ix
 u (pz)-i u (r z)  }\right ) 
 \,  G_Q(x,y;t) \, 
\theta( x+y \leq 1)\,  dx \, dy
  + ``P_Q"{\rm -term}.
\label{312} \end{eqnarray} 
They are given by  the  sum of ``$a+\bar a$'' 
distributions:
\begin{equation}
  F_Q(x,y;t)   =   \sum_{a=1}^{N_f}
(  F_a(x,y;t) + F_{\bar a} (x,y;t) ) \hspace{1cm}
 ;   \hspace{1cm}
 G_Q(x,y;t)   =   \sum_{a=1}^{N_f}
(  G_a(x,y;t) + G_{\bar a} (x,y;t) )
 \, .
\label{313} \end{equation}

\section{Evolution equations}

The QCD perturbative expansion 
for  the matrix element  in Eq.(\ref{21})
generates $\ln z^2$  terms.  As a result,
 limit $z^2 \to 0$ is singular and 
the distributions  $F(x,y;t)$, etc.,
 contain 
logarithmic ultraviolet divergences 
which require an additional 
$R$-operation characterized by some subtraction scale
$\mu$:  $F(x,y;t)\to F(x,y;t\,  |     \,  \mu)$.
The $\mu$-dependence of  $ F(x,y;t\,  |     \,  \mu)$  is governed by the
evolution equation
\begin{equation} 
 \mu \frac{d}{d\mu}  \, F_a(x,y;t\,  |     \,  \mu) =
\int_0^1 \int_0^1 \, \sum_b \,  R^{ab}(x,y;\xi,\eta) \, 
  F_b(\xi,\eta;t\,  |     \,  \mu)\, \theta(\xi +\eta \leq 1) \,
d \xi \, d \eta \,  , 
 \label{41} \end{equation}
where $a,b = G,Q$.
A similar set of equations, with kernels  denoted by 
$\Delta R^{ab}(x,y;\xi,\eta)$  prescribes  the evolution of 
 the parton helicity sensitive 
distributions  $G^a(x,y;t\,  |     \,  \mu)$. 
Since the evolution kernels do not depend on $t$,
from now on we will drop the   $t$-variable from the arguments
of $F(x,y;t\,  |     \, \mu)$ in all cases when this dependence 
is inessential (likewise, the $\mu$-variable  
will be ignored in our notation when it is not important).

Since  integration over $y$ converts $F_a(x,y;t=0\,  |     \, \mu)$ 
into the parton distribution function $f_a(x\,  |     \, \mu)$,
whose evolution is described by the DGLAP equations \cite{gl,ap,d} 
\begin{equation}
\mu \frac{d}{d \mu}  f_a(x\,  |     \,  \mu) =
\int_x^1  P_{ab}(x / \xi;g) 
f_b( \xi\,  |     \,  \mu) \, \frac{d\xi}{\xi}  \,  , 
\label{42} \end{equation}
the kernels $R^{ab}(x,y; \xi, \eta;g)$ must
satisfy the reduction relation
\begin{equation}
\int_0^ {1-x}  R^{ab}(x,y; \xi, \eta;g) \,  d y =
\frac{1}{\xi}\,   P^{ab}(x/\xi;g)\, . 
\label{43} \end{equation}
Aternatively, integration over $x$ converts  $F_a(x,y;t=0\,  |     \, \mu)$ 
into an object similar to a meson distribution amplitude (DA), so 
one may expect that
the result of integration of 
 $R^{ab}(x,y; \xi, \eta;g)$
over $x$ should be related 
to  the    kernels  governing the DA evolution \cite{tmf,bl}.
For the diagonal kernels 
the relations  are rather simple:
\begin{equation}
\int_0^{1-y}   R^{QQ}(x,y; \xi, \eta;g) d x = V^{QQ}(y,\eta;g) \,  
\label{44} \end{equation}
for the quark kernel and a slightly more complicated expression
for the gluon kernel:
\begin{equation}
\int_0^{1-y}  \frac{ x}{ \xi}\, 
 R^{GG}(x,y; \xi, \eta;g) d x = V^{GG}(y,\eta;g) \, .  
\label{45} \end{equation}
The $x/\xi$ factor appears because of the extra  $x$ which was  added 
in the definition  of the gluon DD 
by analogy with the definition for the usual gluon densities.
The nondiagonal kernels  $R^{GQ}$ and $R^{QG}$
obey more complicated reduction formulas (see the Appendix).

The  reduction properties of the diagonal evolution kernels 
can be illustrated using the explicit form of 
the $QQ$-kernel:
\begin{eqnarray}
&& \hspace{-1.5cm} R^{QQ}(x,y;\xi, \eta;g) = 
\frac{\alpha_s}{\pi} C_F \frac1{\xi}
\left \{  \theta  (0 \leq x/\xi \leq 
{\rm min} \{ y/\eta, \bar y / \bar \eta \} ) -
\frac1{2} \delta(1-x/\xi) \delta(y-\eta) \right. \label{46}
 \\
 && \hspace{-1.5cm}  +\left.   
\frac{\theta (0 \leq x/\xi \leq 1) x/\xi}{ (1-x/\xi)} 
\left [ \frac1{\eta}\delta(x/\xi - y/\eta) + 
\frac1{\bar \eta} \delta(x/\xi - \bar y/ \bar \eta) \right]
-2\delta(1-x/\xi) \delta(y-\eta)
\int_0^1 \frac{z}{1-z} \, dz \right \}.
\nonumber
\end{eqnarray}
Here the last (formally divergent) term, as usual,
provides the regularization for the $1/(x-\xi)$  singularities 
present in  the kernel. This singularity 
can be also written as $1/(\eta -y)$  for the term containing $\delta(x/\xi - y/\eta)$ 
and as $1/(\bar \eta - \bar y)$ for the term with $\delta(x/\xi - \bar y/ \bar \eta)$.
Depending on the chosen form
of the singularity,  incorporating   the $1/(1-z)$ term into a plus-type  distribution,
one should treat $z$ as $x/\xi$, $y/\eta$ or $\bar y /\bar \eta$. 
 One can check that  integrating  $R^{QQ}(x,y;\xi, \eta;g)$ 
over $y$ or $x$ gives 
the DGLAP splitting function $P^{QQ}(x/\xi;g)$  and  the 
DA   evolution kernel 
$V^{QQ}(y, \eta;g)$ , respectively: 
\begin{eqnarray}
&& P^{QQ}(z;g) = \frac{\alpha_s}{\pi} 
C_F \left (\frac{1+z^2}{1-z} \right )_+ , 
\label{47} \\
&& V^{QQ}(y, \eta; g)  = \frac{\alpha_s}{\pi} C_F 
\left \{  \left (\frac{y}{\eta} \right )
\left [1+ \frac1{ \eta -y} \right ]
\theta(y \leq \eta) + 
\left (\frac{\bar y}{\bar \eta} \right )
\left [1+ \frac1{y- \eta} \right ]
\theta(y \geq \eta) 
  \right \}_+  \,  . 
\label{48} \end{eqnarray}
Here, ``+'' denotes the standard ``plus''
regularization \cite{ap}.

A convenient  way to get  explicit expressions 
for $R^{ab}(x,y; \xi, \eta;g)$  is 
to extract them from the  kernels $B^{ab}( u,v )$
describing  the 
evolution equations  for the  light-ray operators 
\cite{gdr,bb,bgr,gevol}
\begin{equation}
 \mu \frac{d}{d \mu} \, 
{\cal O}_a(0,z)    =
\int_0^1  \int_0^{1}  
\sum_{b} B^{ab}(u,v ) {\cal O}_b( uz, \bar vz) \,
\theta (u+v \leq 1) \, du \, d v  \,  . 
\label{49} \end{equation}

Since  the definitions of the gluon distributions  
$ F^g(x,y;t)$, $G^g (x,y;t)$
contain an extra $(pz)$ factor on the right-hand side,
which results in  the 
 differentiation $\partial / \partial x$ 
of the relevant kernel, 
it is convenient to proceed in two steps.
 First, we introduce the auxiliary kernels
$r^{ab}(x,y;\xi,\eta;g)$  directly related  by 
\begin{eqnarray}
&&  r^{ab}(x,y; \xi, \eta;g) = 
\int_0^1 \int_0^1 \delta (x- \xi (1-u-v)) \, 
\delta (y-u -\eta (1-u-v)) \, B^{ab} (u,v)\,
 \theta (u+v \leq 1)\, du  \, dv  \nonumber \\
&& \hspace{3cm}= \frac1{\xi} 
B^{ab} ( y - \eta x/\xi, \bar y - \bar \eta x/\xi)
\label{410} \end{eqnarray}
to  the    light-ray evolution 
 kernels $B^{ab}(u,v )$.
The second step is to get the  $R$-kernels 
using the  relations 
\begin{eqnarray}
R^{QQ}(x,y;\xi, \eta;g) = r^{QQ}(x,y;\xi, \eta;g) \  \    ,   \  \ 
R^{GG}(x,y;\xi, \eta;g) = \frac{\xi}{x} \, r^{GG}(x,y;\xi, \eta;g) 
\label{78} \\ 
\frac{\partial}{\partial x} \biggl (x R^{GQ}(x,y;\xi, \eta;g) \biggr ) =
-  r^{GQ}(x,y;\xi, \eta;g)
\,  \ \  , \  \
R^{QG}(x,y;\xi, \eta;g) = - \xi \, \frac{\partial}{\partial x} 
\,r^{QG}(x,y;\xi, \eta;g) \,  . 
 \label{420} \end{eqnarray}
Hence, to obtain  $R^{GQ}(x,y;\xi, \eta;g)$,  we should integrate 
$r^{GQ}(x,y;\xi, \eta;g)$ with respect to $x$.
We  fix  the integration ambiguity  by the requirement
that  $R^{GQ}(x,y;\xi, \eta;g)$ vanishes  for $x>1$.
Then 
\begin{equation}
R^{GQ}(x,y;\xi, \eta;g) = \frac1{x} \, \int_x^1 r^{GQ}(\tilde x,y;\xi, \eta;g)
\, d \tilde x  \,  . 
\label{421} \end{equation}
This convention guarantees a  simple relation  (\ref{43}) 
to  the DGLAP kernels. 
Explicit expressions for the  evolution kernels and discussion 
of evolution equations in the singlet case is given in the Appendix
(see also ref. \cite{bgrdouble}).

\section{Parton interpretation 
and models for double distributions}

\begin{figure}[htb]
\mbox{
   \epsfxsize=10cm
 \epsfysize=5cm
 \hspace{3.5cm}  
  \epsffile{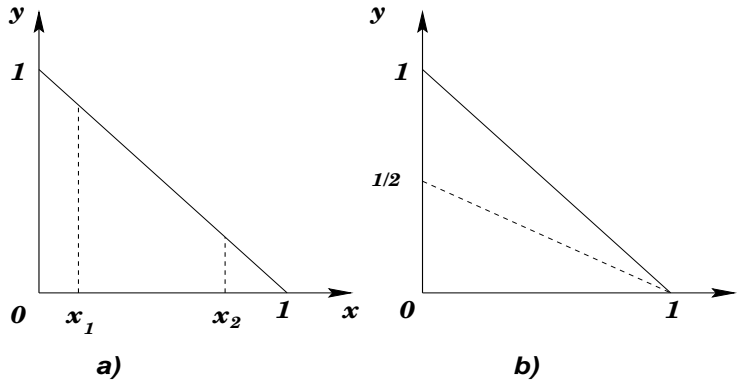}  }
  \vspace{0.5cm}
{\caption{\label{fig:1} $a)$ Integration lines 
in the $(x,y)$-plane giving  reduction 
of  double distributions
$F(x,y;t=0)$ to usual parton densities 
$f(x_1)$ and $f(x_2)$.  $b)$  Symmetry line $y= (1-x)/2$  
for double distributions.
   }}
\end{figure}

The structure of the integrals relating 
double distributions with the usual ones 
\begin{equation}
f_{a,\bar a,g}(x) = \int_0^{1-x} F_{a,\bar a,g}(x,y) dy 
 \end{equation} 
[where $F(x,y)\equiv  F(x,y;t=0$) ] 
 has a simple graphical    illustration
(see Fig.\ref{fig:1}a). 
The DDs  $F(x,y)$ live  on the triangle
defined by $0 \leq x,y,x+y \leq 1$.
 Integrating  $F(x,y)$
over a line parallel to the $y$-axis, we get $f(x)$.
The  reduction formulas and the  interpretation of
the  $x$-variable of 
$F(x,y)$ as a fraction of 
the  $p^+$  momentum 
suggests that the   profile of $F(x,y)$ 
in the  $x$-direction is basically driven by the shape 
of $f(x)$.
On the other hand, the profile in the $y$-direction  
characterizes the spread of momentum induced by
the momentum transfer $r^+$. Hence, 
the $y$-dependence  of $F(x,y)$ 
for fixed $x$ should be 
similar  to that of a distribution 
amplitude $\varphi(y)$.
By analogy with, e.g.,  the pion 
distribution amplitude $\varphi_{\pi}(y)$,
which is symmetric with respect to the change
$y \leftrightarrow 1-y$,
one may expect that the distribution 
of the $r$-momentum between 
the two   partons  described by the same field
should also have some symmetry. 
However, the symmetry cannot 
be as simple as $y \leftrightarrow 1-y$
since the initial $p$ and the 
final $p' \equiv p-r$  momenta are not treated symmetrically
in our description: the variable $x$ 
specifies the fraction of the {\it initial} momentum
$p$ both for the outgoing ($xp+yr$) and incoming 
($xp-(1-y)r$) partons.  To treat $p$ and 
$p'$ symmetrically, we should interpret  $x$
for the returning  parton as the 
fraction of the final hadron momentum 
$p'=p-r$, i.e., rewrite its momentum $xp-(1-y)r$ as
$x(p-r)-(1-x-y)r$. 
Hence, the symmetry  of 
 a  double distribution $F(x,y)$ may be  only  with respect
to the interchange $y \leftrightarrow 1-x-y$
\cite{lech}.

Another way to make the symmetry between the initial and final 
hadrons more explicit is to use 
$P \equiv (p+p')/2$ and $r$ as the basic 
momenta rather than $p$ and $r$ (cf. \cite{ji,ji2,lech})
writing the  momenta of the partons as
$  x P + \tilde y r$ and $  x P - (1- \tilde y) r$.
Then the $y \leftrightarrow 1-x-y$ symmetry 
corresponds to $ \tilde y \leftrightarrow 1- \tilde y$ symmetry.
The variable $\tilde  y$ changes in the interval
$x/2 \leq \tilde y \leq (1-x/2)$.
Writing  
$\tilde y$ as   $\tilde y =(1+ \alpha)/2$,
we introduce a new variable $\alpha$
satisfying a symmetric constraint 
$- \bar x \leq \alpha \leq \bar x$, where 
$\bar x \equiv 1-x$. 
The $y \leftrightarrow 1-x-y$ symmetry 
now converts into  $\alpha \leftrightarrow - \alpha$ symmetry.
Finally,  rescaling $\alpha$ as $\alpha = \bar x \beta$
produces the  variable $\beta$ with $x$-independent limits:
$-1 \leq \beta \leq 1$.  Written in terms of $x$ and $\beta$,
a modified double distribution $\tilde F(x, \beta)$ obeys the  reduction
formula 
\begin{equation}
\frac{\bar x}{2} \int_{-1}^1 \tilde F(x,\beta) d\beta= f(x) \, .
 \end{equation}
 It is instructive to study   some simple  models 
allowing to satisfy this relation. Namely, let us assume 
that the profile in $\beta$-direction is  
 a  universal function  $g(\beta)$ for all $x$, i.e.,  take
 the  factorized ansatz
 \begin{equation}
\tilde F(x,\beta) =  \frac{2}{1-x} \, f(x) \, g(\beta)  \, ,
 \end{equation}
with $g(\beta)$ normalized by 
\begin{equation}
 \int_{-1}^1 g(\beta)\,  d \beta = 1 \, .
 \end{equation}
Possible simple choices for  $g(\beta)$ may be $\delta(\beta)$
(no spread in $\beta$-direction), $\frac34(1-\beta^2)$
(characteristic shape for asymptotic limit 
of quark distribution amplitudes), $\frac{15}{16}(1-\beta^2)^2$
(asymptotic shape of gluon distribution amplitudes), etc.
In our original variables $x,y$, the 
factorized ansatz can be written as
\begin{equation}
F(x,y) =  \frac{h(x,y)}{h(x)}\,  f(x)  \, , \label{65}
 \end{equation}
where $h(x,y)$ is a function 
symmetric with respect to the interchange 
$y \leftrightarrow 1-x-y$.
A trivial observation is  that the variable 
 $x$ itself is given by a combination $[1-(1-x-y)-y]$
symmetric with respect to 
the $y \leftrightarrow 1-x-y$ transformation.
The normalization 
function $h(x)$ is specified by
\begin{equation}
h(x) =  \int_0^{1-x} h(x,y) \, dy \, .
 \end{equation}

 For the three simple choices mentioned above,
the model (\ref{65}) gives 
\begin{equation}
F^{(0)} (x,y) =  \delta(y - \bar x /2) \, f(x)  \ ,  \ 
F^{(1)}(x,y) = \frac{6 y (1-x-y)}{(1-x)^3} \, f(x) \ ,  \
F^{(2)}(x,y) = \frac{30 y^2 (1-x-y)^2}{(1-x)^5} \, f(x) \,  .
 \end{equation}
In a similar way, one can construct ans\"atze 
for functions $F(x,y;t)$ involving  nonzero $t$ values.

\section{Relation to nonforward distributions}

\begin{figure}[htb]
\mbox{
   \epsfxsize=17cm
 \epsfysize=4.5cm
 \hspace{1cm}  
  \epsffile{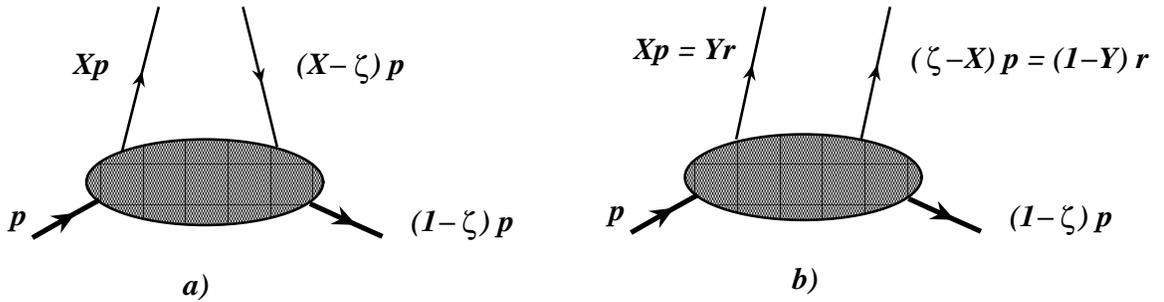}  }
{\caption{\label{fig:nonfwd}   Parton interpretation  
of  nonforward distributions. $a)$ Region $X>\zeta$.
$b)$ Region $X<\zeta$.
   }}
\end{figure}

 The nonforward matrix elements  
accumulate   process-independent information and, hence,
have a quite general nature.
The coefficient of proportionality between 
$p^+$ and $r^+$  characterizes
the skewedness of matrix elements.
The characteristic   feature  implied by
 representations for double distributions 
 (see, e.g., Eqs.(\ref{31}), (\ref{32}))
 is the absence 
of the $\zeta$-dependence in 
the DDs   $F(x,y)$ and $G(x,y)$.
An alternative way to parametrize 
nonforward matrix elements of light-cone operators
is to use   the ratio 
 $\zeta = r^+/p^+$ and the total 
 momentum fraction $X \equiv x+y \zeta$ 
  as  independent 
variables.  Taking into account 
that for a lightcone dominated process 
only one direction for $z$ gives the leading contribution, 
one can do the change $(rz)=\zeta (pz)$  directly
in our definitions of  double distributions.
As a result, the  variable $y$ would appear there  
 only in the 
$x+y\zeta \equiv X$ 
combination,
where $X$  can be treated as    the {\it total}  fraction 
of the initial hadron momentum $p$ carried by the active  quark.
If we require that 
the light-cone plus component 
of the final hadron momentum (i.e., $p^+ - r^+$)
is positive, then  $0 \leq \zeta \leq 1$.
Using the spectral property   $0 \leq x+y \leq 1$
of  double distributions we obtain that 
the variable $X$ satisfies a similar ``parton''   
constraint $0\leq X \leq 1$.
Integrating each particular  double   distribution 
$F_{a, \bar a, g}(X-y \zeta,y)$ 
over $y$ gives  the nonforward  parton distributions  
\begin{equation}
{\cal F}_{\zeta}^{a, \bar a, g} (X) = \theta(X \geq \zeta) 
 \int_0^{ \bar X / \bar \zeta } F_{a, \bar a, g}(X-y \zeta,y) \, dy + 
 \theta(X \leq \zeta)  \int_0^{ X/\zeta} F_{a, \bar a, g}
 (X-y \zeta,y) \, dy \,  , 
\label{71}  \end{equation}
where $\bar \zeta \equiv 1- \zeta$.
The two components of NFPDs correspond to positive
($X> \zeta$) and negative ($X< \zeta$)
values of the fraction $X' \equiv X - \zeta$ 
associated with the returning parton.
As explained in refs. \cite{gluon,npd},
the second component  can be interpreted as the
probability amplitude for the initial hadron with momentum 
$p$ to split into the final hadron with momentum $(1-\zeta)p$
and the two-parton state with total momentum $r=\zeta p$
shared by the partons
in fractions $Yr$ and $(1-Y)r$, where $Y=X/\zeta$
(see Fig.\ref{fig:nonfwd}).

For the gluon DDs,
 the $y \leftrightarrow 1-x-y$ symmetry 
holds only if, instead of $(z \cdot p)$,  
 one uses the symmetric overall factor $(z \cdot P)$ 
in the definitions 
(\ref{36}),(\ref{37})\footnote{I am grateful to
G. Piller for attracting my attention to this 
point.}. The use of such a definition of the gluon DDs is 
implied in this section. Furthermore,  
the nonforward gluon distribution
${\cal F}_{\zeta}^g (X)$ is obtained by integrating 
$x F_g(x,y)|_{x=X-y \zeta}$. 
To simplify notations, it will be also implied 
below that, 
 for the gluons,    $F(X-y \zeta, y)$ 
in Eq.(\ref{71}) corresponds to $(X-y \zeta) F_g(X-y \zeta, y)$.

\begin{figure}[htb]
\mbox{ \hspace{2cm} 
   \epsfxsize=8cm
 \epsfysize=5cm   \epsffile{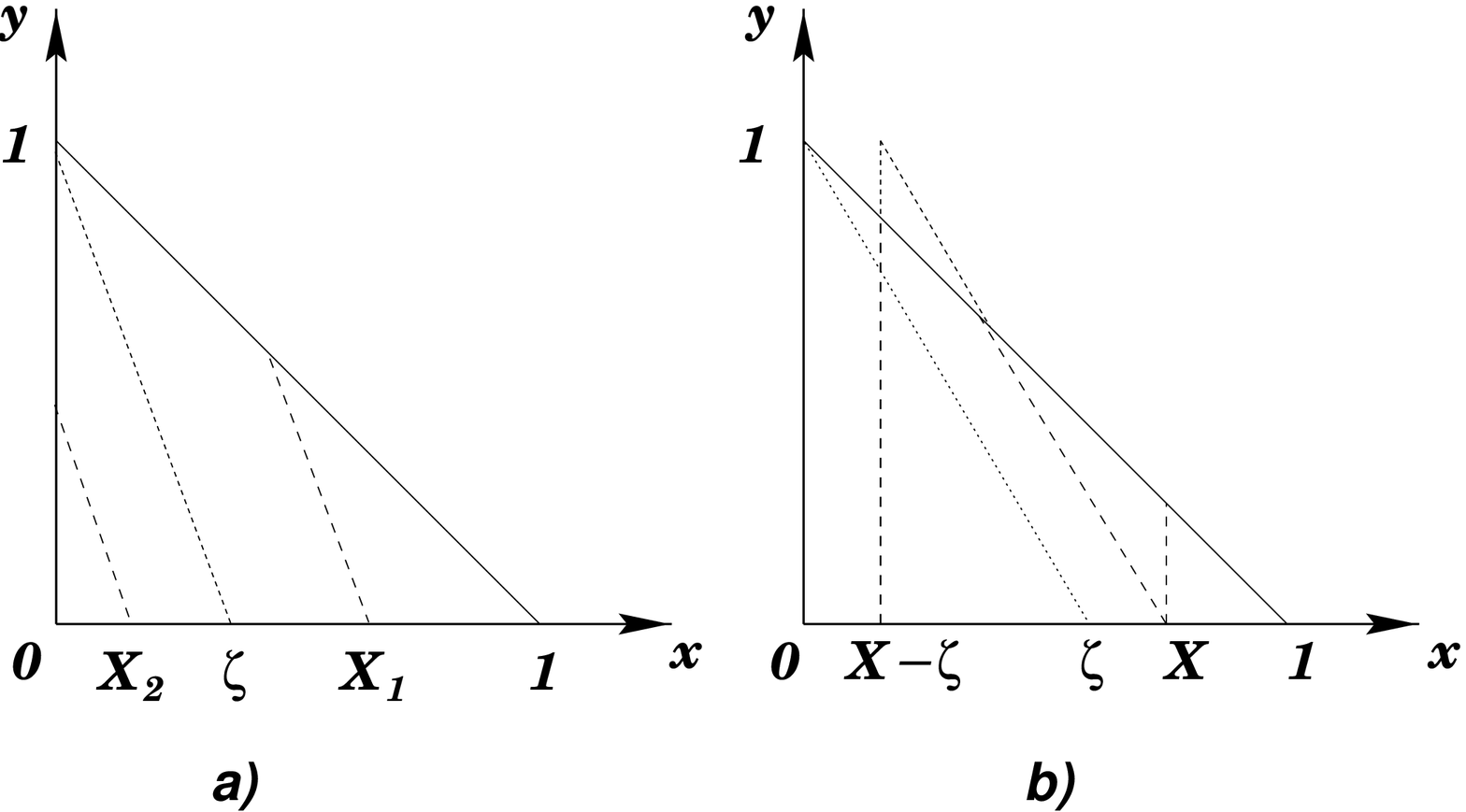} \hspace{0.3cm}
 \epsfxsize=4cm
 \epsfysize=5cm
  \epsffile{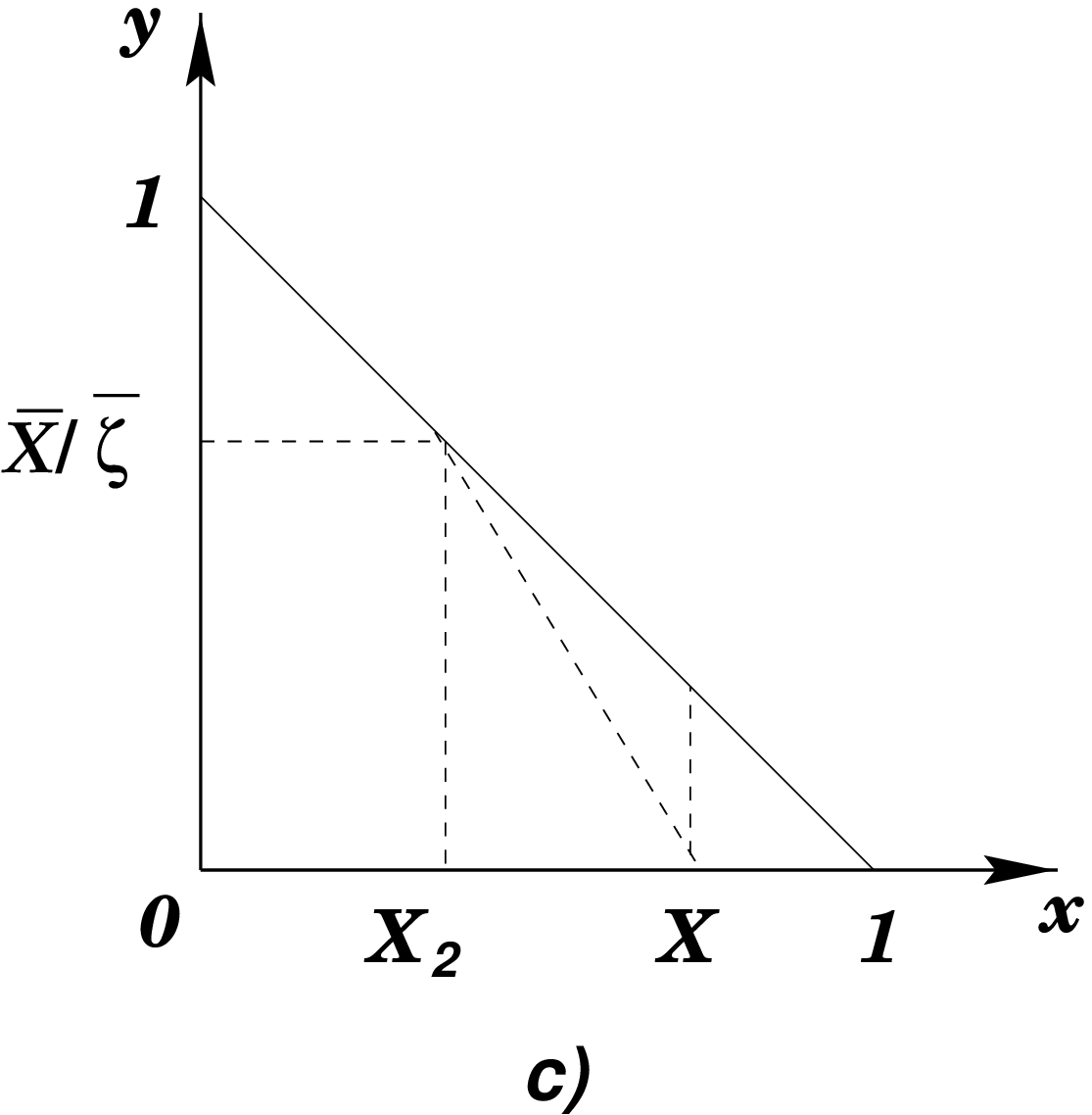}  }
  \vspace{0.5cm}
{\caption{\label{fig:2} Relation between double distributions
$F(x,y)$ and nonforward parton distributions 
${\cal F}_{\zeta}(X)$. $a)$ Integration lines 
for three cases: $X_1 > \zeta$, $X = \zeta$ and $X_2 < \zeta$.  
$b,c)$ Comparison of 
integration lines for the nonforward parton distribution 
${\cal F}_{\zeta}(X)$ and usual parton densities $f(X)$, 
$f(X')$ (shown in  2$b$) and $f(X)$, $f(X_2)$
with $X_2 = X'/\bar \zeta $  (shown in 2$c$).
   }}
\end{figure}

The  basic distinction  between  
  double  distributions $F(x,y)$ and 
nonforward distributions 
${\cal F}_{\zeta} (X)$ is that
 NFPDs  explicitly depend on the skewedness  parameter
$\zeta$. They   form 
 families of functions   ${\cal F}_{\zeta}^{a, \bar a, g}  (X)$ 
whose shape changes when $\zeta$ is changed.
The fact that the functions ${\cal F}_{\zeta} (X)$
corresponding to different $\zeta$'s 
are obtained by integrating  the same 
double distribution $F(x,y)$ 
imposes essential  restrictions on possible 
shapes of ${\cal F}_{\zeta} (X)$ and on how 
they change with
changing $\zeta$.
The relation between NFPDs
and  DDs  has a simple graphical    illustration
on the ``DD-life'' triangle
defined by $0 \leq x,y,x+y \leq 1$ (see Fig.\ref{fig:2} ).
To get ${\cal F}_{\zeta} (X)$,  one should 
integrate $F(x,y)$ over $y$ along a straight
line specified by $x=X- \zeta y$. Fixing some value of  $\zeta$,
one deals with  a set of parallel lines corresponding to 
different values of $X$.
Evidently,  each such  line intersects the $x$-axis
at $x=X$. The upper limit of the $y$-integration
is determined by intersection of this line
either with the line $x+y=1$ (this happens if 
$X > \zeta$)  or with the $y$-axis (if $X < \zeta$).
The line corresponding to $X=\zeta$
separates the triangle into two parts
generating  two components of the
nonforward parton distribution.
In the forward case, when $\zeta =0$,
there is only one component, and the usual
parton densities $f(x)$ are produced
by integrating $F(x,y)$ along the vertical lines $x=$const
(see Fig.\ref{fig:1}).
In case when $X> \zeta$,
looking at the  integration line for the nonforward 
parton distribution ${\cal F}_{\zeta}(X)$
one can see (Fig.\ref{fig:2}b) that it is
inside the space  between the integration lines 
giving   the usual parton densities 
 $f(X)$ and $f(X')$  corresponding to 
the momentum fractions $X$, $X' \equiv X- \zeta$
of the initial and final parton.  
Assuming a monotonic decrease of the double distribution
$F(x,y)$ in the $x$-direction and a universal profile 
in the $y$-direction, one may expect that 
 ${\cal F}_{\zeta}(X)$  is larger than  $f(X)$ but smaller than  $f(X')$.
Inequalities between forward and nonforward distributions
were recently discussed in refs. \cite{maryskin,jirev,pisoter}.
They are based on the application of the Cauchy-Schwartz
inequality 
\begin{equation}
| \sum_S \langle \,  H(p'); X'p, S \,  | \, H(p); Xp, S \, \rangle |^2  \leq  
\sum_S \langle \, H(p), Xp, S \, | \, H(p); Xp,S \, \rangle 
\sum_{S'} \langle \, H(p'); X'p,S' \, | \, H(p'); X'p,S'\,  \rangle \, ,
\end{equation}
to  the nonforward 
distributions ${\cal F}_{\zeta} (X)$ written generically
as $${\cal F}_{\zeta} (X) = 
\sum_S \langle \, H(p'); X'p,S\,  | \, H(p); Xp,S \,  \rangle \, , $$
where  $| \, H(p); Xp,S \,  \rangle$ describes 
 the probability amplitude that the
hadron with momentum $p$  converts  into a parton  with momentum 
$Xp$ and spectators $S$. 
The forward matrix elements are identified with  the usual
parton densities  
\begin{equation}
\sum_S \langle \, H(p); Xp,S \, | \, H(p); Xp,S \,  \rangle \, = f(X).
\end{equation}
Notice  that the hadron momentum in the second 
forward matrix element is $p' = \bar \zeta p$,
hence the argument of the relevant parton density is 
$X'/\bar \zeta$,  and one has 
\begin{equation}
\sum_S \langle \,  H(p'); X'p,S \, | \, H(p'); X'p,S \,  \rangle = 
\sum_S \langle \,  H(p'); X'p'/\bar \zeta,S  \, | \, H(p'); 
X'p'/ \bar \zeta,S  \,  \rangle = 
f(X'/\bar \zeta\, ) /\bar \zeta \, . \label{ineq}
\end{equation}
As a  result, we obtain (compare with 
\cite{jirev,pisoter})
\begin{equation}
{\cal F}_{\zeta}^q(X) \leq \sqrt{f(X)f(X'/\bar \zeta\, )/\bar \zeta } \leq
\frac1{2 \sqrt{1- \zeta} } \, \biggl [f(X)+ f(X'/\bar \zeta\, ) \biggr ]  \, .
\label{basineq} \end{equation}
In other words, the functions involved in the bound
for ${\cal F}_{\zeta}(X)$ are  $f(X)$
and $f(X_2)$ where the fraction $X_2 \equiv  X'/\bar \zeta$ is  
larger  than  $X'$ \cite{jirev,pisoter}. 
One can  see that $X_2$  is given  exactly by the $x$-value
of the intersection point  in which
the integration line $x=X-\zeta y$ giving  the nonforward
distribution ${\cal F}_{\zeta}(X)$ crosses the
 boundary line   $x+y=1$ (see Fig.\ref{fig:2}c).
 For the gluon nonforward distributions, one should take into
account extra factors $(zp)$, $(zp')$ present in the definitions
of forward distributions and the 
overall factor chosen  in the
 rhs of definitions of the nonforward gluon distributions
(see  Eqs. (\ref{36}), (\ref{37})).
 If one uses the $p \leftrightarrow
 p'$ symmetric  combination  $(z \cdot P)$, then 
 \begin{equation}
{\cal F}_{\zeta}^{g \, (symm)} (X) \leq
\frac1{1- \zeta/2} \sqrt{f(X)f(X'/\bar \zeta\, )} \leq
\frac1{2(1-\zeta/2) } \, \biggl [f(X)+ f(X'/\bar \zeta\, ) \biggr ]  \, .
\label{modineq} \end{equation}

It is clear  that the whole construction 
makes sense only if $X' >0$ (or $X > \zeta$).
If $X' < 0$, the nonforward distribution corresponds to  matrix 
elements  $\langle H(p' );  Xp, X'p, S \,   | \, H(p),S \, \rangle $
which have  no obvious relation to the usual parton densities.
Furthermore,  in our  graph of 
Fig.\ref{fig:2}a, the left end of the line 
  $x=X-\zeta y$ in this case corresponds to $x=0$, where the usual
parton densities
are infinite, and the   inequalities become trivial.
In fact, they are  trivial  even for the border 
point $X=\zeta$. 
 Another deficiency of the Cauchy-Schwartz-type inequalities is that
they do not give the lower bound for nonforward 
distributions though our  graphical interpretation
suggests that ${\cal F}_{\zeta}(X)$  for $X>\zeta$ 
is larger than $f(X)$
if the $x$-dependence of the double distribution 
$F(x,y)$ along the lines $y =k \bar x $ is monotonic.

To develop intuition  about possible shapes of 
nonforward distributions, it is instructive to derive 
the  NFPDs corresponding to three simple 
models specified in the previous section.
In particular, for the \mbox{$F^{(0)}(x,y) = \delta (y -\bar x/2)
f(x)$}
ansatz we get 
\begin{equation}
{\cal F}_{\zeta}^{(0)} (X) = \frac{\theta(X \geq \zeta/2)}{1-\zeta/2}
 f \left (\frac{X-\zeta/2}{1-\zeta/2} \right ) \, ,
 \label{model}
  \end{equation}
i.e.,  NFPDs for non-zero  $\zeta$ are obtained from 
the forward distribution $f(X)\equiv {\cal F}_{\zeta} (X)$  
 by  a shift and rescaling. 
Note that the model (\ref{model}) 
satisfies the   inequalities (\ref{basineq}), 
(\ref{modineq})  in the region $X > \zeta$
for any function $f(x \, | \, Q_0)$ 
of  the \mbox{$f(x \, | \, Q_0) = A x^{-a}(1-x)^b $} type 
provided that  $a \geq 0$  and  $b> 0$. 
  Using the relations 
\begin{equation} 
H (x,\xi; t)=  {(1-\zeta/2)} \, {\cal F}_{\zeta} (X;t)  \  \  ;  \  \   
 \tilde x =  \frac{X-\zeta/2}{1-\zeta/2}  \  \  ;  \  \  
\xi = \frac{\zeta}{2- \zeta}   \label{onfpd}
 \end{equation}
between our nonforward distributions and 
  Ji's off-forward parton distributions  (OFPDs) $ H(x,\xi;t)$ \cite{ji,ji2},
  one can see that the delta-function ansatz gives the simplest 
  $\xi$-independent model 
  $$H^{(0)}(x,\xi; t=0) = f(x)$$ for OFPDs at $t=0$
  \footnote{Since hadrons are  massive, $t=0$ is outside the
  physical region; hence, the $t \to 0$
  limit should be understood 
  in the sense of analytic continuation.}. 
It is worth noting that the MIT bag model calculation
\cite{jimong} did produce a set of OFPDs 
which are almost independent of $\xi$.
An evident interpretation is  that
the model constructed 
in ref.\cite{jimong} strongly suppresses the redistribution 
of the momentum transfer among the constituents which results in  
a very narrow spread of $F(x,y)$   in the $y$-direction.
Even if such a picture is physically
correct for a low normalization point $Q_0 \sim 500 \,$ MeV,
evolution to higher values $Q \gtrsim 1 \, $ GeV 
widens the $y$-profile of $F(x,y)$ and evolved  
OFPDs would change their shape with $\xi$, as was explicitly 
demonstrated through  a  numerical calculation 
by Belitsky {\it et al.} \cite{bel}.

\begin{figure}[htb]
\mbox{
   \epsfxsize=8cm
 \epsfysize=5cm
 \hspace{4cm}  
  \epsffile{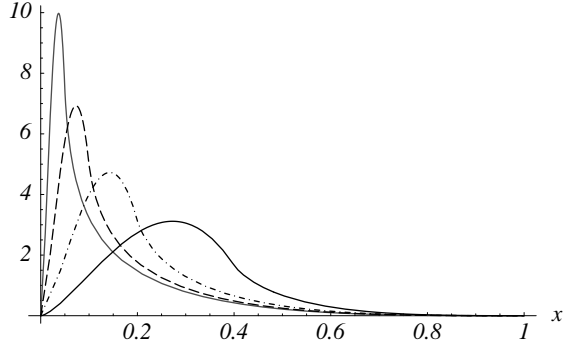}  }
  \vspace{0.5cm}
{\caption{\label{fig:3} Nonforward parton distributions 
$ {\cal F}_{\zeta}^{(1)} (X)$ for different values of the skewedness
$\zeta=0.05$ (thin line), $\zeta=0.1$ (dashed line),
$\zeta=0.2$ (dash-dotted line) and $\zeta=0.4$ (full line)
in the ``valence quark oriented''  model specified by Eq.(6.11) for $a=0.5$.  
   }}
\end{figure}

The evolution of {\it nonforward } 
distributions ${\cal F}_{\zeta}(X\,  |     \, Q)$ 
was recently studied in refs.  \cite{ffgs,maryskin,gokmar}. 
As a starting condition, the authors   assume  that,
at some low scale $Q_0$,  the nonforward
distributions 
$ {\cal F}_{\zeta}(X\,  |     \, Q)$ for all $\zeta$ 
have the same  universal shape 
 coinciding with that
of the usual (forward)  densities $f(X,Q_0)$.
This assumption 
corresponds to the ansatz 
 $F(x,y\,  |     \, Q_0) = \delta(y) f(x\,  |     \, Q_0)$
with double distribution being nonzero on 
the $x$-axis only.  
This ansatz is not realistic,
since it has no symmetry with respect
to the $y \leftrightarrow (1-x-y)$ interchange.
However, evolution equations are  applicable
to any distribution and, just due to its  asymmetric 
profile,
this unrealistic double  distribution has 
a very distinctive   evolution pattern
reflecting the restoration of the $y \leftrightarrow 1-x-y$
symmetry. 
Namely, 
the asymptotic functions $F(x,y\,  |     \, Q \to \infty)$ 
are  
$y \leftrightarrow (1-x-y)$ symmetric. In particular, both
in pure gluodynamics and in QCD,  we have  
 $F_g(x,y\,  |     \, Q \to \infty) \sim y^2 (1-x-y)^2$
 (see Ref.\cite{gluon} and the Appendix).
 Hence, 
one may expect that the evolition 
of $F_g(x,y\,  |     \, Q ) $ 
 shifts its
crest  towards the $y= \bar x/2$ line and also 
makes the $y$-shape of
the  double distribution wider.
To see whether  the results of 
Refs.\cite{ffgs,maryskin} reflect  
this expectation,
 we  
introduce  a   general model with a narrow
$y$-dependence: $F^{(0)}_k (x,y) = \delta (y- k \bar x)f(x)$
(in what follows, it will be referred to as the ``$k$-delta ansatz'').
This double distribution is concentrated on the $y=k \bar x$ line 
and gives 
 \begin{equation}
 {\cal F}_{\zeta}^{(0)(k)} (X) = \frac1{1-k\zeta} \, 
f \left ( \frac{X-k \zeta}{1-k\zeta} \right ) 
 \end{equation}
for nonforward distributions. 
In case of  two other models, simple  analytic results 
can be obtained only if we specify a model for $f(x)$.
For the ``valence quark''-oriented ansatz $F^{(1)}(x,y)$,
the following choice of a normalized distribution
\begin{equation} f^{(1)}(x) = 
 \frac{\Gamma(5-a)}{6 \,  \Gamma(1-a)}
  x^{-a} (1-x)^3  \label{74} \end{equation}
is   $(a)$ close 
to phenomenological  valence quark distributions
and $(b)$  produces a simple expression
for the double distribution since the denominator
$(1-x)^3$ factor in Eq.(\ref{74}) is canceled.
As a result, the integral in Eq.(\ref{71})
is easily performed and  we get
\begin{equation}
{\cal F}_{\zeta}^{(1)} (X) =\frac{4-a}{\zeta^3}
\left \{  X^{2-a} (\zeta \bar a \bar X - 2 (X-\zeta))
+\theta(X \geq \zeta)  \left( \frac{X-\zeta}{1-\zeta} \right )^{2-a}
(\zeta \bar a \bar X + 2 X\bar \zeta) 
 \right \} \, .
 \end{equation}
 Resulting curves for $ {\cal F}_{\zeta}^{(1)} (X)$ 
with $a=0.5$ and   $\zeta = 0.05,
 0.1, 0.2, 0.4$ are shown in Fig.\ref{fig:3}.
 A characteristic feature of each curve 
is a  maximum located close to 
 the relevant border 
 point $X = \zeta$ and slightly shifted 
 to the left from it. 
Note that both the functions  $ {\cal F}_{\zeta}^{(1)} (X)$ 
and their derivatives $ (d/dX) {\cal F}_{\zeta}^{(1)} (X)$
are continuous at $X=\zeta$.
The latter property is secured by the fact that
$F^{(1)}(x,y)$ vanishes at the upper corner $x=0,y=1$.
The $(1-x)^5$ denominator factor 
for the ``gluon-oriented'' ansatz $F^{(2)}(x,y) $ is canceled if 
one takes the model 
 $f(x)  \sim x^{-a} (1-x)^5$ which, fortunately, is also 
consistent with the $x \to 1$ behavior of the 
phenomenological gluon distributions. 
It is   well known \cite{compton,gluon} 
that the values of nonforward distributions
$  {\cal F}_{\zeta} (X)$ taken at the border point $X=\zeta$
determine imaginary parts of DVCS and hard electroproduction 
amplitudes. 
 An interesting question 
is the relation between the usual distributions 
$f(\zeta)$ and the  values 
$  {\cal F}_{\zeta}(\zeta)$
of nonforward distributions at the border point.
It is easy to calculate that for the $k=1/2$
delta ansatz $F^{(0)}(x,y)$
this ratio is  given by 
\begin{equation}
    R^{(0)}(\zeta)\equiv {\cal F}_{\zeta}^{(0)}(\zeta)/f(\zeta) 
    =\frac{f(\zeta/(2-\zeta))}{(1-\zeta/2)f(\zeta)}  \,.   
    \end{equation}
It is larger than 1 
 for any monotonically descreasing function $f(x)$,
i.e., the  nonforward distribution $  {\cal F}_{\zeta}(\zeta)$
in this case  is larger than $f(\zeta)$. 
In the small-$\zeta$ limit, $R^{(0)}(\zeta)$ 
is completely determined by
the small-$x$ behavior of $f(x)$, and  the 
expression for  $R^{(0)}(\zeta) $ simplifies to 
\begin{equation}
    R^{(0)}(\zeta)\,  |     \, _{ \zeta \ll 1} 
\approx \frac{f(\zeta/2)}{f(\zeta)}  \, .  
    \end{equation}
Hence, if $f(x)$ has a purely powerlike
behavior  $f(x) \sim x^{-a}$ for small
$x$, then   $R^{(0)}(\zeta \to 0) =2^a (1+ O(\zeta))$,
i.e., for small $\zeta$, the ratio of the nonforward  distribution
 $  {\cal F}_{\zeta}(\zeta)$  and the usual
parton density $f(\zeta)$ 
 is practically constant, deviating from the 
$\zeta =0$ limiting value by $O(\zeta)$ terms only.
The limiting value in this case  is 1.41 
for $a=0.5$ and 1.23 (1.15) for $a=0.3$ ($a=0.2$).
However, if $f(x)$ is a sum of two different powerlike terms
 $ A x^{-a}+B x^{-b}$
 or if it contains logarithms, e.g.,
 $f(x) \sim x^{-a} \ln(1/x)$ for small $x$,
then the $\zeta$-dependence  is more pronounced.
In  the latter case
\begin{equation}
    R^{(0)}(\zeta)  
   \approx  2^a \left (1 + \frac {
\ln 2 }{\ln(1/\zeta)} \right ) \,  ,  
    \end{equation}
and there is a visible deviation from 
the limiting $\zeta \to 0$ value for all 
accessible $\zeta$: on the $\ln(1/\zeta)$ scale, 
the $\zeta$-dependence  
of the ratio  $ R^{(0)}(\zeta) $ cannot be neglected 
even for $\zeta \sim  10^{-5}$.

For a  general $k$-delta  model 
$F^{(0)}_k (x,y) = \delta (y- k \bar x)f(x)$,
the ratio  $  {\cal F}_{\zeta}(\zeta)/f(\zeta)$  
for small $\zeta$ can be approximated
by $f(\zeta(1-k))/f(\zeta)$ which again gives
a $\zeta$-independent constant $(1-k)^{-a}$
for a purely powerlike function $f(x) \sim x^{-a}$
while the $\ln(1/x)$-factor would modify the constant
by $[1+\ln(1-k)/\ln \zeta]$.

If one uses the ``valence quark''-oriented ansatz 
$F^{(1)}(x,y)$ with a simple powerlike
behavior $f(x) \sim x^{-a}$ for small $x$,
the  ratio is  given by 
  \begin{equation}
    R^{(1)}(\zeta)\equiv {\cal F}_{\zeta}^{(1)}(\zeta)/f(\zeta) 
    = \frac1{(1-\zeta)^2(1-a/2)(1-a/3)} \, . 
    \end{equation}
Just like in the previous example,
 the  nonforward distribution $  {\cal F}_{\zeta}(\zeta)$
is  larger than $f(\zeta)$ for all positive $a$.
For small $\zeta$, the ratio tends to $1/(1-a/2)(1-a/3)$,
e.g., to $1.6$ for $a=0.5$ which is the usual choice
for valence quark distributions (for comparison,
taking  $a=0.3$ ($a=0.4$) gives $1.3$ (1.44) for 
$R^{(1)}(\zeta\to 0) $). For small $a$, this result
can be translated into 
   $ R^{(0)}(\zeta \to 0)  
    \approx   e^{5a/6}  \approx f( e^{5/6} \zeta)/f(\zeta)$,
which coincides with the ratio  $ R^{(0)}(\zeta \to 0) $
for the modified narrow ansatz 
$F^{(0)}_k (x,y) = \delta(y- k \bar x) f(x)$ with $k \approx 0.56$.
Hence, for $ {\cal F}_{\zeta}(\zeta)$
the widening of  the $y$-distribution 
can be approximated by a  narrow distribution shifted 
from $y=\bar x/ 2$ upwards to the $y \approx k \bar x$ line.
Again, a  logarithm $\ln(1/x)$ in $f(x)$ at small
$x$ would induce a visible $\zeta$-dependence
for  the $ R^{(1)}(\zeta)  $ ratio even for very small
$\zeta$.

Switching to  the ``gluon-oriented'' ansatz   
$F^{(2)}(x,y)$ with a purely
power behavior $f(x) \sim x^{-a}$
for small $x$,  we obtain  a similar expression 
\begin{equation}
    R^{(2)} (\zeta) \equiv {\cal F}_{\zeta}^{(2)}(\zeta)/f(\zeta) = 
    \frac{1}{(1-\zeta)^3(1-a/3)(1-a/4)(1-a/5)}  \, ,   
    \end{equation}
which is close to $(2.17)^a$
for small $a$.
 To approximate this result
  by the delta ansatz  $F^{(0)}_k (x,y)$,
one should take $k \approx 0.54$. The effective shift
upward is smaller in this case because $F^{(2)}(x,y)$
is more narrow in the $y$-direction than $F^{(1)}(x,y)$.

Choosing $a$, we should take
into account that the nonforward gluon distribution
$ {\cal F}_{\zeta}^{g}(X)$ reduces to $Xf_g(X)$ in the 
$\zeta \to 0$ limit \cite{gluon,npd}. Hence,
$f(\zeta)$ in the above formulas should be understood
as $\zeta f_g(\zeta)$. Now, if we make an 
old-fashioned assumption
that  $Xf_g(X)$ tends to a constant
as $X\to 0$, then $a=0$ and $R^{(2)}(\zeta)$
tends to 1 at small $\zeta$, i.e., the nonforward 
distribution ${\cal F}_{\zeta}^{g}(\zeta)$ 
coincides in the small-$\zeta$ limit 
with its forward counterpart
$\zeta f_g(\zeta)$. 
To get a more realistic 
 gluon distribution  $Xf_g(X)$ growing at small  $X$ 
one should use a positive parameter $a$. 
Taking $a=0.3$, we get  $R^{(2)}(\zeta \to 0) \approx 1.27$, 
and $R^{(2)}(\zeta \to 0) \approx 1.17$  (1.39) 
for $a=0.2$ ($a=0.4$)).

These estimates  for the  ratio 
${\cal F}_{\zeta}^{g}(X \,  |     \, Q)/Xf_g(X\,  |     \, Q)$ 
are close to those 
 obtained in refs. \cite{ffgs,maryskin} 
where    the  nonforward distributions 
$ {\cal F}_{\zeta}^g(X\,  |     \, Q)$
at high normalization point $Q$ 
were constructed by applying evolution  
equations to an initial low normalization point $Q_0$ ansatz  
$ {\cal F}_{\zeta}^g(X\,  |     \, Q_0)$
which was assumed to have a universal $\zeta$-independent shape 
coinciding with the usual distribution
$Xf_g(X\,  |     \, Q_0)$.
In particular, Martin and Ryskin
considered the evolution of the  
gluon NFPD  in pure gluodynamics. They  took 
$Q_0^2 = 1.5$ GeV$^2$ (two other choices
$Q_0^2 = 0.4$ GeV$^2$ and $Q_0^2 = 4$ GeV$^2$
were also considered) and then 
evolved $ {\cal F}_{\zeta}^g(X\,  |     \, Q)$
to higher $Q^2$ values $Q^2 = 4, 20$, and  100 GeV$^2$.
They  found that   $R(10^{-5} ) \approx 1.3$ 
for $Q^2 =100$ GeV$^2$, which corresponds to
$a \approx 0.3$ in   our  $F^{(2)}$ model.
This value   is close
to those  used in phenomenological 
parametrizations of the gluon distributions.
It should be also noted that the results for $R(\zeta)$ 
obtained in  Ref.\cite{maryskin} have a nonnegligible 
$\zeta$-dependence. 
This feature can be expected since  
the GRV gluon distribution \cite{GRV}
which they use can be rather well approximated 
at $Q^2= 4 $ GeV$^2$ by a simple formula 
$$xf_g^{GRV}(x,Q^2 = 4 \, {\rm GeV}^2)
\approx \frac14 \, x^{-0.3} \ln(1/x)$$ which works with
$10 \%$ accuracy for $x$ ranging from $10^{-1}$ 
to $10^{-5}$.  In the pure gluodynamics
approximation used in Ref.\cite{maryskin}, 
its shape does not drastically
change when  evolved either to $Q^2 = 1.5$ GeV$^2$ or 
to $Q^2 = 20$ and 100 GeV$^2$.

\begin{figure}[htb]
\mbox{
   \epsfxsize=3.5cm
 \epsfysize=2.5cm
  \epsffile{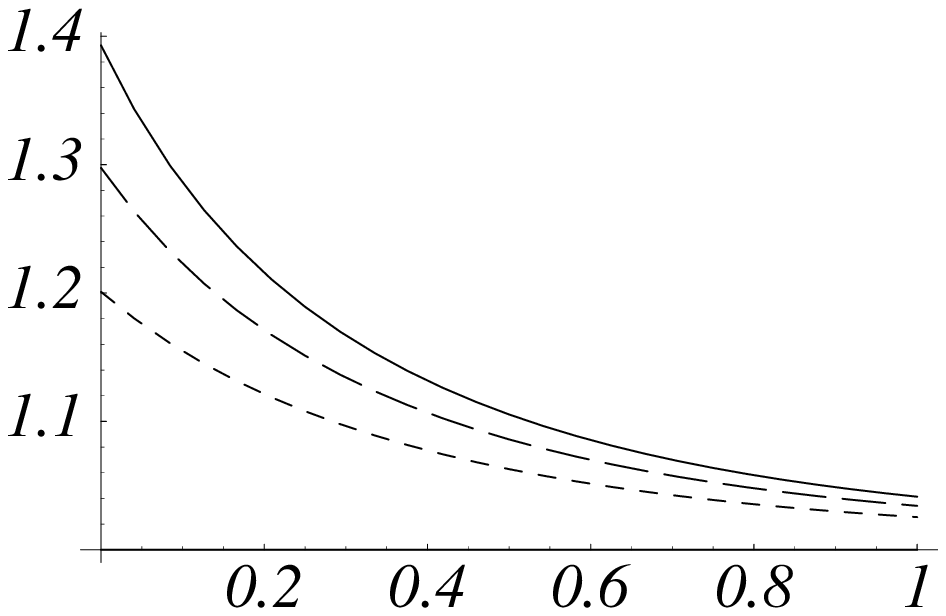} 
  \epsfxsize=3.5cm
 \epsfysize=2.5cm  \hspace{0.5cm}  \epsffile{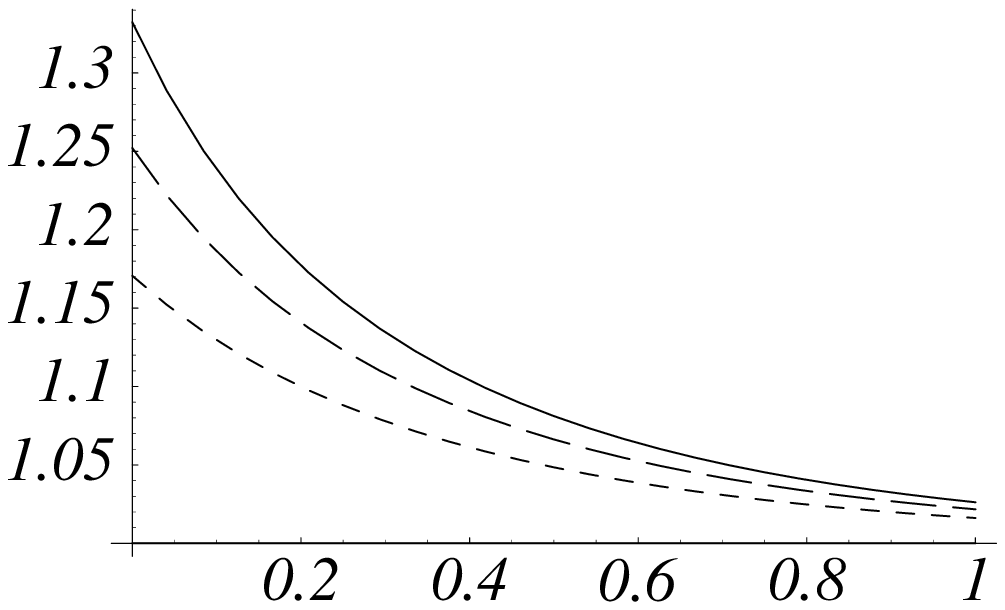}
  \epsfxsize=3.5cm
 \epsfysize=2.5cm \hspace{0.5cm} \epsffile{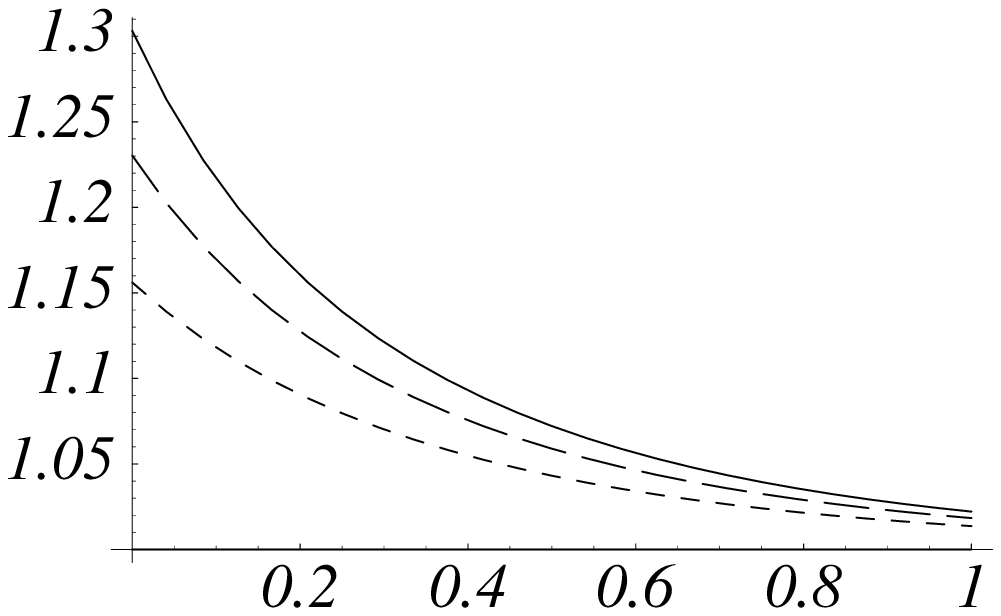}
  \epsfxsize=3.5cm
 \epsfysize=2.5cm \hspace{0.5cm} \epsffile{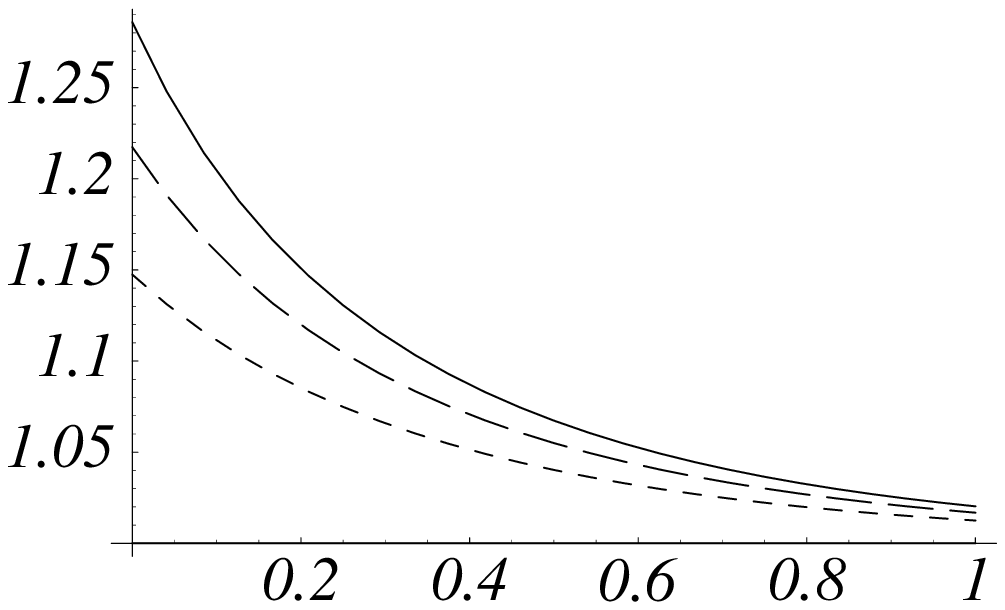}
   }
\\ . \hspace{2cm} $a)$  \hspace{3.5cm} $b)$  
\hspace{3.5cm} $c)$ \hspace{3.5cm} $d)$ 
  \vspace{0.5cm}
{\caption{\label{fig:mr}  Ratio ${\cal F}_{\zeta}^g(X) / Xf_g(X)$
vs. $\log_{10} (X/\zeta)$ 
as obtained from the model given by Eq. (6.16) 
for $a)$ $ \zeta = 10^{-2}, b)$ $  \zeta = 
10^{-3}, c) $ $ \zeta = 10^{-4}$ and $d)$ $  
\zeta = 10^{-5}$ with $k=0.48$ (solid lines),
$k=0.4$ (long-dashed lines) and $k=0.30$ (short-dashed lines). 
    }}
\end{figure}

As discussed above, the assumption that the 
nonforward distributions $ {\cal F}_{\zeta}^g(X\,  |     \, Q_0)$
have  a universal $\zeta$-independent shape
corresponds to the   ansatz 
 $F^{(0)}_k(x,y\,  |     \, Q_0) = \delta (y) \,  f (x \, |\, Q_0)$,
i.e., to the $k$-delta ansatz with the 
 vanishing slope $k=0$. 
Modeling the evolved double distributions 
by a   $k$-delta  ansatz with nonzero $k$, we  expect 
that, due to the restoration of the
$y \to 1-x-y$ symmetry,
  the effective slope parameter $k$
should increase with $Q^2$.
Namely, for the  $k$-delta ansatz,  
the ratio of the nonforward distribution
${\cal F}_{\zeta}^g(X)$ and the forward  parton
distribution $f(x) \equiv Xf_g(X)$ is given by 
\begin{equation}
R(X, \zeta) \equiv \frac{ {\cal F}_{\zeta}^g(X) }{Xf_g(X)}
=  \frac{ f(X-k \zeta \bar \zeta/(1- k \zeta))}
{(1- k \zeta ) f(X)} \, . 
 \end{equation}
Taking $f(x) = \frac14 \, x^{-0.3} \ln (1/x)$
and the $Q^2$-dependent slope $k(Q^2) = 0.3; 0.4 ; 0.48$ for 
$Q^2 = 4 ; 20 $ and 100 GeV$^2$, respectively,
we were able to reproduce the 
results of Ref.\cite{maryskin} for a wide range 
of $\zeta$ parameters: $\zeta = 10^{-2},
10^{-3}, 10^{-4}$ and $10^{-5}$. 
The relevant curves, coinciding with those of Ref. \cite{maryskin} 
 within a few per cent accuracy, 
 are shown in Fig.\ref{fig:mr}.
Hence,  the increase of the ratio $R(X, \zeta)$ 
with $Q^2$ observed in Refs.\cite{maryskin,ffgs}
basically reflects the 
 shift of 
the gluon double distribution from the
$x$ axis $y=0$ towards the symmetry line
$y = \bar x /2$.  This effect, being an  artifact 
of the initial conditions, 
 plays the dominant role
up to $Q^2 \sim 100$ GeV$^2$. 
As argued above, the $\zeta$-dependence 
of the  ratio may be traced to the fact
that the gluon distribution $xf_g^{GRV}(x)$ 
differs from a simple power $x^{-a}$.

Since  the assumption   ${\cal F}_{\zeta} (X \, | \, Q_0)
 = f(X  \, | \, Q_0)$
is equivalent to  the ansatz 
 $F(x,y\,  |     \, Q_0) = \delta(y) 
 f(x\,  |     \, Q_0)$ which is not 
symmetric with respect to the 
$y \to (1-x-y)$ interchange,
 one should avoid using
it  as a starting  condition for evolution. 
As explained earlier, a more 
realistic set of nonforward
distributions
\begin{equation}
{\cal F}_{\zeta}^{(0)} (X\,  |     \, Q_0) = 
\frac{\theta(X \geq \zeta/2)}{1-\zeta/2}
 f \left (\frac{X-\zeta/2}{1-\zeta/2} \right ) \label{712}
  \end{equation}
is generated by the  
$F^{(0)}(x,y\,  |     \, Q_0)=\delta(y-\bar x/2) f(x)$
ansatz for the double distribution
corresponding to skewedness-independent set
of Ji's off-forward distributions.
Comparing these two sets, one may be tempted to argue     
that for extremely small $\zeta$ 
considered in Ref.\cite{maryskin}, 
  $\zeta/2$ terms in Eq.(\ref{712}) are  inessential.
Of course, $\zeta/2$ can be neglected 
when subtracted from  1.  However, 
for the $X$-values close to the
border point $X =\zeta$, the shift by
$\zeta/2$    produces visible changes for functions
having the  $X^{-a}$ behavior with $a \sim 0.3$.
In the case of the ansatz (\ref{712}),
the ratio  $R(X,\zeta \, | \,Q ) \equiv
{\cal F}_{\zeta}^g(X \, | \, Q) / Xf_g(X \, | \,Q )$ 
differs from 1 for all $Q$. 
For small $\zeta$, the difference 
is significant only for $X$ close to $\zeta$.

When  a narrow  double distribution has its crest 
on the $y = \bar x/2$ line from the very start,
there are no effects due to the shift of the crest,
and  the  $Q$-evolution of $R(X,\zeta \, | \,Q )$ 
in the region $X > \zeta$   reflects only the widening 
of the double distribution   in the $y$-direction
and the change of its  profile in the $x$-direction.
As we have seen,  the widening of the 
double distribution   changes the effective
slope $k$ by a small amount only.
Hence, for small $\zeta$ one  can use the 
approximate formula  
\begin{equation}
{\cal F}_{\zeta}^g(X \, | \, Q) |_{\zeta \ll 1} 
\approx   (X- \zeta/2) f_g(X- \zeta/2 \, | \,Q ) \label{eq:617}
  \end{equation}
 for evolved distributions as well.
In other words, the ratio 
$R(X, \zeta \, | \,Q )$  for $X> \zeta$ and small
$\zeta$ 
can be estimated from existing results 
for the usual gluon density $f(X) \equiv X f_g(X \, | \,Q )$.

Comparing the formula (\ref{eq:617})  with the 
relation (\ref{onfpd}) between our nonforward and Ji's off-forward
distributions, one can conclude that Eq.(\ref{eq:617}) 
is equivalent to a statement
 that at small $\xi$ and $\tilde x > \xi$ 
one can neglect the $\xi$-dependence  of the 
off-forward distributions $H(\tilde x;\xi)$.
Again, such a statement is only nontrivial 
if $\tilde x \sim \xi$.
To analyze the accuracy of eq.(\ref{eq:617}), 
we will construct an expansion of $H(\tilde x;\xi)$ 
in powers of $\xi$.
To this end, it is convenient to use the parton picture 
based on modified double distribution 
$\tilde F(x, \alpha )$  in which 
the plus component of the parton momenta
is  measured in units of that of the average hadron momentum 
$P=(p+p')/2$.  
The parton momenta then are 
$xP + (1+\alpha )  r/2$ and $xP - (1-\alpha   )r/2$ 
with $\alpha  $ changing between $ -\bar x $ and  $\bar x$.
Defining $r^+ / P^+ = 2 \xi$ and $\tilde x = x + \xi \alpha$, 
one obtains the description 
in terms of  the off-forward parton distributions $H(\tilde x;\xi)$ 
\cite{ji,ji2}. The parton  momenta are now
$(\tilde x + \xi) P $ and  $(\tilde x -\xi) P $. In the
region $\tilde x > \xi$, the OFPDs 
are obtained from $\tilde F(x, \alpha  )$  by the 
integral
\begin{equation}
H(\tilde x;\xi)|_{\tilde x  > \xi} = 
\int_{- (1- \tilde x)/(1+\xi)}^{(1- \tilde x)/(1-\xi)}
\tilde F(\tilde x - \xi \alpha   ,\alpha  ) \, d  \alpha   \, .
\label{offf} 
\end{equation} 
 Using the $\alpha \to - \alpha$ symmetry of $\tilde F (x,\alpha)$, 
it is easy to see from this expression that the  
 off-forward parton distributions
 $H(\tilde x;\xi)$  are even functions of $\xi$:
\begin{equation}
H(\tilde x;\xi) = H(\tilde x;- \xi)  \, .
\end{equation} 
This result was originally obtained by X. Ji \cite{jirev} 
with the help of a different technique.
Expanding the rhs of Eq.(\ref{offf}) in powers of $\xi$,
we get
\begin{equation}
H(\tilde x;\xi) = f(\tilde x) +  
\xi^2 \left [ 
\frac12 \int_{- (1- \tilde x)}^{(1- \tilde x)}
\frac{ \partial^{2} \tilde F(\tilde x,  \alpha  )}{\partial \tilde x^{2}}
 \,\alpha  ^{2} \, d  \alpha   \,  + (1- \tilde x)^2
\left. \left (\frac{ \partial \tilde F (\tilde x,  \alpha  )}
 {\partial \alpha } 
 -2 \frac{ \partial \tilde F (\tilde x,  \alpha  )}
 {\partial \tilde x  } \right ) \right |_{\alpha = 1 - \tilde x}
 \right ] +  \ldots \, .
\label{exxi2} 
\end{equation} 
where  $f(\tilde x)$ is  the forward distribution.
Hence, for small $\xi$, the corrections 
are formally $O(\xi^2)$, i.e., they look very small.
However,   if $ \tilde F ( x,  \alpha  )$  
has a singular 
behavior like $x^{-a}$, then 
$$\frac{ \partial^{2}\tilde F(\tilde x,  \alpha  )}{\partial \tilde x^{2}}
\sim \frac{a (1+a)}{\tilde x^2} F(\tilde x,  \alpha  )$$
and the relative suppression of the first correction 
is $O(\xi^2/\tilde x^2)$  i.e., the corrections 
are tiny for all $\tilde x$ except for the region  $\tilde x \sim \xi$
 where the correction has   no parametric smallness.
Nevertheless, even in this region it is suppressed 
numerically, because the $\alpha  ^2$ moment is rather small
for   a distribution concentrated in the small-$\alpha  $ region.
 This discussion shows that the formula (\ref{eq:617}) is not just an
automatic  consequence of the
$O(\xi^2)$ nature of the first nonvanishing correction.
It is easy to write expicitly all  the   terms 
 which are not suppressed in the $\tilde x \sim \xi \to 0$ limit
\begin{equation}
H(\tilde x;\xi) = \sum_{k=0}^{\infty} \frac{\xi^{2k}}{(2k)!} 
\int_{- (1- \tilde x)}^{(1- \tilde x)}
\frac{ \partial^{2k}\tilde F(\tilde x,  
\alpha  )}{\partial \tilde x^{2k}}
 \,\alpha  ^{2k} \, d  \alpha   \,  + \ldots  \  .
\label{expanxi} 
\end{equation} 
 The numerical suppression of higher terms is even stronger,
 and the series converges rather fast.

 In terms of the off-forward distributions,
 the inequality (\ref{basineq}) reads
 \begin{equation}
 H^q(x,\xi)  \leq \sqrt{ \frac1{1-\xi^2} \,
 f \left ( \frac{x+\xi}{1+\xi} \right )
 f \left ( \frac{x-\xi}{1-\xi} \right )} \leq 
  \frac1{2 \sqrt{1-\xi^2}} \left  [f \left ( \frac{x+\xi}{1+\xi} \right )
+  f \left ( \frac{x-\xi}{1-\xi} \right ) \right ]\, .
\label{Hq}
 \end{equation}
 For the gluons,   one should use  the
 inequality  (\ref{modineq}), which
 leads to
 \begin{equation}
 H^g(x,\xi)  \leq \sqrt{ 
 f \left ( \frac{x+\xi}{1+\xi} \right )
 f \left ( \frac{x-\xi}{1-\xi} \right )} \leq 
  \frac1{2 } \left  [f \left ( \frac{x+\xi}{1+\xi} \right )
+  f \left ( \frac{x-\xi}{1-\xi} \right ) \right ]\, .
\label{Hg}
 \end{equation}
 Again, if one takes  the model $H(\tilde x, \xi) = f(\tilde x)$, the
inequalities (\ref{Hq}) and (\ref{Hg})  are valid for any function $f(x)$ of
$x^{-a}(1-x)^b$ type  with $a \geq 0, \, b>0$.

So far we assumed in our models that DDs are 
finite everywhere on the ``life triangle''.
Consider, however, a situation when the partons 
 emerge from a
meson-like state (or glueball/pomeron in the 
gluon case) exchanged in the  $t$ channel. 
In this case, the partons  just share 
the plus component of the momentum transfer $r$:
information about 
the magnitude of the  initial hadron momentum
is lost if the exchanged particle can be described
by a pole propagator $\sim 1/(t-m_M^2)$.
Hence, the meson-exchange  contribution to a double distribution
is proportional to $\delta(x)$ or its derivatives, e.g.:
\begin{equation}
F^M (x,y) \sim  \delta (x) \, \frac{\varphi_M (y)}{m_M^2 -t }
\end{equation} 
 where $\varphi_M (y)$ is the distribution amplitude of the meson $M$.
This contribution to the nonforward 
distribution  is nonzero
only in the $0 <X< \zeta$ region:
\begin{equation} {\cal  F}_{\zeta}^M (X)
\sim   \frac{\varphi_M (X/\zeta)}{\zeta (m_M^2 -t )} \, \theta (0 \leq X \leq \zeta).
\end{equation} 
At the beginning, we described the nonforward matrix element
of a quark operator by two functions $F^a(x,y)$ and  $F^{\bar a}(x,y)$
corresponding to positive-$x$ and negative-$x$ parts of the
general Fourier representation. Since $x=0$ for a meson-exchange 
contribution, it makes sense to treat it 
as a third independent  component, i.e., to parametrize the
nonforward matrix element by the sum 
$F^a \oplus F^{\bar a} \oplus  F^{M}$.
All three components  contribute to the 
nonforward distributions in the
$0 \leq X \leq \zeta$ region.
However, the $\delta (x)$ terms do not contribute 
to the nonforward distributions in the
$  X \geq \zeta$ region and to the  usual parton densities $f(x)$.
For this reason, the $\delta (x)$ terms, if they exist, 
would lead to violation of sum rules 
(like energy-momentum sum rule) for the usual parton 
densities.

Note that if the meson DA $\varphi (y)$ does not vanish at the end-points,
the  nonforward distribution does not vanish at $X=0$
(the off-forward parton distributions $H(\tilde x ; \xi)$
in this case are discontinuous at $x = \pm \, \xi$).
As explained in ref. \cite{npd},  pQCD factorization for DVCS and other
hard electroproduction processes fails in such a situation,
because of the  $1/X$ factors 
($1/(\tilde x \pm \xi)$ factors if OFPD formalism is used)
contained in  hard amplitudes.  
It should be mentioned that a nearly discontinuous behavior
of OFPDs for $\tilde x = \pm \, \xi$ 
was obtained in the
chiral soliton model \cite{ppp}.
Formally, the evolution to sufficiently high $\mu$
results in  the functions vanishing at
the end-point $X=0$. A non-trivial question, however,
is whether evolution starts at all in a 
situation when pQCD factorization fails.

\section{Summary}

In this paper, we duscussed the formalism 
of double distributions. We treated them as 
the starting objects in parametrization of 
nonforward matrix elements. 
An alternative description in terms of 
nonforward or off-forward parton distributions
was  obtained by an appropriate integration
of the relevant DDs.
Incorporating spectral and symmetry properties
of double distributions, we proposed 
simple models producing self-consistent sets
of non-forward distributions ${\cal F}_{\zeta}(X)$ 
and discussed their $\zeta$-dependence and
relation to usual (forward) parton densities.
Using a qualitative picture of the evolution
of  double distributions, we were able to explain
and model  the basic features of the evolution pattern 
of nonforward  distributions observed 
in numerical evolution studies \cite{maryskin}.
In the Appendix, we present the set of evolution
equations for double distributions in the singlet case
and  discuss their analytic solution. 
Work on   numerical  
evolution of the nonforward distributions
corresponding to  realistic ans\"atze (\ref{712}) 
is  in progress \cite{progress}.
Another interesting problem for a future investigation
is  a  numerical  evolution of double distributions.

\section{Acknowledgements}

I acknowledge stimulating discussions  and communication 
with I. Balitsky,  A. Belitsky, J. Blumlein, V. Braun,
J. Collins, C. Coriano, L. Frankfurt,
 B. Geyer, K. Golec-Biernat, X. Ji, L. Mankiewicz, 
 I. Musatov, G. Piller, M. Polyakov,
 D. Robaschik, M. Ryskin,  M. Strikman and O.V. Teryaev. 
 This work was supported by the US 
 Department of Energy under contract
DE-AC05-84ER40150.

\begin{appendix}

\section{Evolution equations for the singlet case}

As described in Sec. IV, 
the  evolution kernels  for 
double distributions can be conveniently obtained
 from  the light-ray evolution kernels
$B^{ab}(u,v)$. For the  parton helicity averaged case, the latter 
were originally obtained in Refs. 
\cite{gdr,bb}. Here we present them in the form
given in Ref.\cite{gluon}:
\begin{eqnarray}
&& 
B^{QQ}(u,v ) = \frac{\alpha_s}{\pi} C_F \left 
(1 + \delta( u) [\bar v/v]_+  + 
\delta(v) [\bar u/ u]_+ - \frac1{2} \delta( u)\delta(v) \right ) \, ,
\\
&&   B^{GQ}(u,v ) = \frac{\alpha_s}{\pi} C_F \biggl 
(2 + \delta( u)\delta(v) \biggr  ) \, ,
\\ 
&&  B^{QG}(u,v ) = \frac{\alpha_s}{\pi} N_f \left 
(1 + 4uv -u -v \right ) \, , \\ 
&&  B^{GG}(u,v ) = \frac{\alpha_s}{\pi} N_c \biggl (
4(1 + 3uv -u -v) + \frac{\beta_0}{2 N_c} \,
\delta(u)\delta(v) 
+ \left  \{ \delta(u) \biggl[ \frac{\bar v^2}{v} - \delta (v) \int_0^1
\frac{d \tilde v}{\tilde v} \biggr ] 
+ \{ u \leftrightarrow v \} \right \} 
 \biggr  )  \, .
\label{414} \end{eqnarray}
As usual,  $\beta_0 = 11- \frac2{3} N_f$ 
is the lowest coefficient of the 
QCD $\beta$-function.
Evolution kernels for the parton helicity-sensitive
case are given by \cite{bgr,gevol} 
\begin{eqnarray} 
&&
\Delta B^{QQ}(u,v ) = B^{QQ}(u,v )
\\
&&  \Delta B^{GQ}(u,v ) = \frac{\alpha_s}{\pi} \, C_F \biggl 
( \delta( u)\delta(v) - 2 \biggr  ) \, ,
\\ 
&&   \Delta B^{QG}(u,v ) = \frac{\alpha_s}{\pi} \, N_f \left 
( 1- u - v  \right ) \, , \\ 
&&  \Delta B^{GG}(u,v ) = B^{GG}(u,v ) - 
12 \,  \frac{\alpha_s}{\pi} \, N_c \, uv .
 \label{418} \end{eqnarray}

At one loop,  $\Delta R^{QQ}(x,y;\xi, \eta;g) = R^{QQ}(x,y;\xi, \eta;g)$,
and this  kernel was 
 already displayed in Eq. (\ref{46}).
Other  kernels, including the  $R^{GG}(x,y;\xi, \eta;g)$ kernel  originally
obtained in Ref.\cite{gluon}, are given by 
\begin{eqnarray}
&&  \Delta R^{GG}(x,y;\xi, \eta;g) = 
\frac{\alpha_s}{\pi} N_c \frac1{\xi} 
\Biggl  \{
4 \, \theta  (0 \leq x/\xi \leq 
{\rm min} \{ y/\eta, \bar y / \bar \eta \} )+
\delta(1-x/\xi) \delta(y-\eta)
\frac{\beta_0}{2 N_c} 
  \label{422}   \\
 && \hspace{1cm}  +  
\frac{\, \theta (0 \leq x/\xi \leq 1) (x/\xi)}{ (1-x/\xi)} 
\left [ \frac1{\eta} \, \delta(x/\xi - y/\eta) + 
\frac1{\bar \eta} \, \delta(x/\xi - \bar y/ \bar \eta) \right]
-2 \delta(1-x/\xi) \delta(y-\eta)
 \int_0^1 
\frac{d \tilde v }{1-\tilde v } \, \Biggr \}\, , 
\nonumber  \\
&&   R^{GG}(x,y ; \xi, \eta ; g) = \Delta R^{GG}(x,y;\xi, \eta;g)
+ 12 \,  \frac{\alpha_s}{\pi} \, N_c \, \frac{1}{x} 
\,  (y-\eta x/\xi)(\bar y - 
\bar \eta x/\xi)  \, \theta  (0 \leq x/\xi \leq 
{\rm min} \{ y/\eta, \bar y / \bar \eta \} ) \, , \\ && 
\Delta R^{GQ}(x,y ; \xi, \eta ; g) = \frac{\alpha_s}{\pi} C_F \, \frac1{x} \, 
\biggl \{  -2 \biggl [ \left ( \frac{y}{\eta} - \frac{x}{ \xi} \right) 
  \, \theta (x / \xi \leq y/\eta \leq 1) 
+ \left \{ {y \to \bar y} \atop { \eta \to \bar \eta} \right  \} \biggr  ] 
+ \delta(\eta -y) \, \theta (0 \leq x \leq  \xi ) \biggr  \}  \, , \\
&&
R^{GQ}(x,y ; \xi, \eta ; g) = \frac{\alpha_s}{\pi} C_F \, \frac1{x} \, 
\biggl \{  2 \biggl  [  \left ( \frac{y}{\eta} - \frac{x}{ \xi} \right) 
\, \theta (x / \xi \leq y/\eta \leq 1) 
+ \left \{ {y \to \bar y} \atop { \eta \to \bar \eta} \right  \} \biggr  ] 
+ \delta(\eta -y) \, \theta (0 \leq x \leq \xi )  \biggr  \}  \, , \\
&&
\Delta R^{QG}(x,y ; \xi, \eta ; g) = \frac{\alpha_s}{\pi} N_f \frac1{\xi} 
\biggl \{ \frac{x}{\xi} \biggl (  \delta(x/\xi - y/\eta) 
\, \theta(y \leq \eta) + \delta(x/\xi - \bar y/ \bar \eta)
\, \theta(y \geq \eta) \biggr )  \nonumber \\ && 
\hspace{9cm} -  \,  
\theta (0 \leq x/\xi \leq {\rm min} \{ y/\eta, \bar y/ \bar \eta  \})
\biggr \}  \, ,  \\ &&
R^{QG}(x,y ; \xi, \eta ; g) = \Delta R^{QG}(x,y ; \xi, \eta ; g)
+4 \, \frac{\alpha_s}{\pi} N_f
\, \frac1{\xi} \, \eta  \bar \eta
\biggl  (  \frac {y}{\eta} + \frac{ \bar y }{ \bar \eta}
- 2\, \frac{x}{\xi} \biggr ) 
\, \theta (0 \leq x/\xi \leq {\rm min} \{ y/\eta, \bar y/ \bar \eta  \}) 
 \,  .
\label{428} \end{eqnarray}

To find a formal solution of  the evolution equations 
for double distributions,  
we proposed in Refs. \cite{compton,gluon}  to combine
the standard methods used to  solve the 
evolution equations for parton densities and distribution amplitudes.
Hence, let us 
 start with taking  the moments with respect to $x$.
Utilizing 
the property 
 $R^{ab}(x,y; \xi, \eta;g) = R^{ab}(x/\xi,y; 1, \eta;g)/\xi$
we get  
\begin{equation}
\mu \frac{d}{d \mu}  F_n^a(y\,  |     \,  \mu) =
\sum_{b} \int_0^1  R_n^{ab} (y,\eta;g) F_n^b(\eta\,  |     \,  \mu)
 \,  d \eta \,  , 
\label{eq:fnev}
 \end{equation}
where  $F_n^a(y\,  |     \,  \mu)$ is the $n$th $x$-moment of $F^a(x,y\,  |     \,  \mu)$
\begin{equation}
F_n^a(y\,  |     \,  \mu) = \int_0^{1} x^n  F^a(x,y\,  |     \,  \mu) dx \, .
\label{eq:fnmom}
 \end{equation}
The kernels $R_n^{ab} (y,\eta;g)$ and analogous 
kernels  $\Delta R_n^{ab} (y,\eta;g)$ 
governing the evolution of $G_n^a(y\,  |     \,  \mu)$
are given by
\begin{eqnarray}
&& R_n^{QQ} (y,\eta;g) = \Delta R_n^{QQ} (y,\eta;g) = 
\frac{\alpha_s}{\pi} C_F 
\left \{  \left (\frac{y}{\eta} \right )^{n+1} 
\left [\frac1{n+1}+ \frac1{ \eta -y} \right ]
\theta(y \leq \eta)  \right. \nonumber \\ && + 
\left (\frac{\bar y}{\bar \eta} \right )^{n+1} 
\left [\frac1{n+1} + \frac1{y- \eta} \right ]
\theta(y \geq \eta) 
\left. - \frac1{2} \, \delta(y-\eta) -  2 \delta(y-\eta)
\int_0^1 \frac{z}{1-z} \, dz  \right \}.
\label{eq:rnkernelQQ}
 \end{eqnarray}
\begin{eqnarray}
&& \Delta R_{n}^{GG} (y,\eta;g) = 
\frac{\alpha_s}{\pi} N_c 
\left \{   \left (\frac{y}{\eta} \right )^{n+1} 
\left ( \frac4{n+1} + 
 \frac{1}{\eta -y} \right ) \theta(y \leq \eta)
+ \left \{  {y \to \bar y} \atop {\eta \to \bar \eta} \right  \}
 + \delta(y-\eta) 
\left [ \frac{\beta_0}{2 N_c} - 
2 \int_0^1 \frac{dz}{1-z} \,  \right ]
 \right \},
\label{eq:rnkernelDGG}
 \end{eqnarray}
\begin{eqnarray}
&& R_{n}^{GG} (y,\eta;g) = \Delta R_{n}^{GG} (y,\eta;g) + 
12 \frac{\alpha_s}{\pi} N_c \frac1{n+1} 
\left \{   \left (\frac{y}{\eta} \right )^{n+1} 
\left ( \frac{ \eta \bar y}{n} -
 \frac{y \bar \eta}{n+2} 
  \right ) \theta(y \leq \eta)
+ \left \{  {y \to \bar y} \atop {\eta \to \bar \eta} \right  \}
 \right. \biggr \},
\label{eq:rnkernelGG}
 \end{eqnarray}
\begin{eqnarray}
&& \Delta R_{n}^{QG} (y,\eta;g) = 
\frac{\alpha_s}{\pi} N_f 
\frac{n}{n+1} \left \{   \left (\frac{y}{\eta} \right )^{n+1} 
\theta(y \leq \eta) + \left (\frac{\bar y}{\bar \eta} \right )^{n+1} 
\theta(y \geq \eta) \right \},
\label{56}
 \end{eqnarray}
\begin{eqnarray}
&& R_{n}^{QG} (y,\eta;g) = \Delta R_{n}^{QG} (y,\eta;g)
+ 4 \,\frac{\alpha_s}{\pi} N_f \, \frac{n}{n+1}
\biggl  \{ \left (\frac{y}{\eta} \right )^{n+1} 
\left ( \frac{ \eta \bar y}{n} 
- \frac{y \bar \eta } {n+2}
  \right ) \theta(y \leq \eta) 
+ \left \{  {y \to \bar y} \atop {\eta \to \bar \eta} \right  \} \biggr  \}
\label{57}
 \end{eqnarray}
\begin{eqnarray}
&& \Delta R_{n}^{GQ} (y,\eta;g) = 
\frac{\alpha_s}{\pi} C_F 
\frac1{n}  \left \{  \delta(y-\eta) 
 - \frac2{n+1}  \left [ \left(\frac{y}{\eta} \right )^{n+1} 
\theta(y \leq \eta) +  \left ( \frac{\bar y}{\bar \eta} \right )^{n+1} 
\theta(y \geq \eta) \right ] \right \},
\label{58}
 \end{eqnarray}
\begin{eqnarray}
&&  R_{n}^{GQ} (y,\eta;g) = 
\frac{\alpha_s}{\pi} C_F 
\frac1{n}  \left \{  \delta(y-\eta) 
 + \frac2{n+1}  \left [ \left(\frac{y}{\eta} \right )^{n+1} 
\theta(y \leq \eta) +  \left ( \frac{\bar y}{\bar \eta} \right )^{n+1} 
\theta(y \geq \eta) \right ] \right \} \, . 
\label{59}
 \end{eqnarray}

From Eqs. (\ref{56}), (\ref{58}) one can derive the following
reduction formulas  for the nondiagonal kernels:
\begin{equation} 
 \frac{\partial}{\partial y}\Delta  R_1^{QG}(y,\eta;g)
 = -  \Delta V^{QG}(y,\eta;g) \, , 
 \end{equation}
\begin{equation} 
\lim_{n \to 0} n  \Delta R_n^{GQ}(y,\eta;g) = - \frac{\partial}{\partial y} 
 \Delta V^{GQ}(y,\eta;g) \,  . 
 \end{equation}
The same relations connect the nondiagonal kernels
 $ R^{GQ},
 R^{QG}$ with the exclusive  kernels 
$ V^{GQ}(y,\eta;g),
 V^{QG}(y,\eta;g)$ given in ref. \cite{npd}. 
 To understand their structure, one should realize
that  
constructing the nondiagonal $QG$ and $GQ$  
kernels,  one faces 
 mismatching $(zp)$ factors 
which in the pure $\zeta=1$ case 
are converted into  derivatives with respect to $y$.

It is straightforward to check 
that  all the kernels $R_n^{a,b} (y,\eta;g)$ 
 (and $\Delta R_n^{a,b} (y,\eta;g)$) 
have the property
$$R_n^{ab} (y,\eta;g) w_n(\eta) =
R_n^{ab} (\eta,y;g) w_n(y) ,$$ where $w_n(y)= (y \bar y)^{n+1}$. 
Hence, the  eigenfunctions  of the evolution equations 
 are orthogonal with
the weight $w_n(y)= (y \bar y)^{n+1}$, $i.e.,$ 
they are proportional to the Gegenbauer polynomials
$C^{n+3/2}_k(y-\bar y)$, see \cite{bl,mikhrad} 
 and 
Refs.\cite{chase,shvys,ohrn,gro} where
the general algorithm was  applied to the 
evolution of flavor-singlet distribution amplitudes.

Expanding  the moment functions  $F_n^{a}(y\,  |     \,  \mu)$
 over the  Gegenbauer polynomials
$C^{n+3/2}_k(y-\bar y)$ 
\begin{equation} 
F_n^{a}(y\,  |     \,  \mu) = (y \bar y)^{n+1} 
\sum_{k=0}^{\infty} F_{nk}^a(\mu)
C^{n+3/2}_k(y-\bar y) 
\label{eq:rneq}
 \end{equation}
we get the evolution equation for the expansion coefficients
\begin{equation}
\mu \frac{d}{d \mu}  F_{nk}^a(\mu) =  \frac{\alpha_s}{\pi}
\sum_{b} \Gamma_{n k}^{ab}  F_{nk}^b(\mu)
  \,  , 
\label{a27}
 \end{equation}
where  $\Gamma_{n k}^{ab}  $ are 
the eigenvalues of the kernels $ R_n^{ab} (y,\eta;g)$
related to the elements $\gamma_{N}^{ab} $
of the usual flavor-singlet anomalous dimension matrix 
\begin{equation}
\Gamma_{n k}^{QQ}  = \gamma_{n+k}^{QQ} \  ,  \  
\Gamma_{n k}^{QG}  =n  \gamma_{n+k}^{QG}   \  ,  \  
\Gamma_{n k}^{GQ}  =\frac{1}{n} \gamma_{n+k}^{GQ}  \  ,  \  
\Gamma_{n k}^{GG}  = \gamma_{n+k}^{GG}  \, ; 
 \end{equation} 
and similarly for the helicity-sensitive 
quantities $\Delta \Gamma_{n k}^{ab}$. 
Namely, 
\begin{eqnarray}
&& \gamma_N^{QQ}= \Delta   \gamma_N^{QQ}= - C_F \left [ \frac12 - 
\frac1{(N+1)(N+2)} +2 \sum_{j=2}^{N+1} \frac1{j}
\right ] \,  , \\ && 
\gamma_N^{GG}= - 2 N_c \left [- \frac1{N(N+1)}  
- \frac1{(N+2)(N+3)}
 + \sum_{j=1}^{N+1} \frac1{j}\right ] 
 + \frac{\beta_0}{2} \,  , \\  && 
\Delta \gamma_N^{GG}= - 2 N_c \left [  
- \frac2{(N+2)(N+3)}
 + \sum_{j=1}^{N+1} \frac1{j}\right ] 
 + \frac{\beta_0}{2} \,  , \\  && 
\gamma_N^{GQ}= C_F \frac{N^2+3N+4}{(N+1)(N+2)}  \hspace{0.5cm} , 
 \hspace{0.5cm} \gamma_N^{QG}= N_f \frac{N^2+3N+4}{N(N+1)(N+2)(N+3)} 
 \,  , \\  &&  
\Delta \gamma_N^{GQ}= C_F \frac{N(N+3)}{(N+1)(N+2)}  \hspace{0.5cm} , 
 \hspace{0.5cm} \Delta \gamma_N^{QG}= N_f \frac{1}{(N+2)(N+3)} 
.
 \label{eq:ads}
 \end{eqnarray}

Let us consider first two simplified situations.
In the quark nonsinglet case, the evolution is governed 
(in helicity-averaged case) by 
$\gamma_{n+k}^{QQ}$ alone:
\begin{equation} 
\tilde F_n^{NS}(y\,  |     \,  \mu) = 
(y \bar y)^{n+1} \sum_{k=0}^{\infty} A^{nk}
C^{n+3/2}_k(y-\bar y) \left [\log (\mu /\Lambda) \right]^
{2\gamma_{n+k}^{QQ}/\beta_0},
\label{eq:fnnons}
 \end{equation}
Since $\gamma_0^{QQ} =0$ while all the  anomalous dimensions
$\gamma_{N}^{QQ}$ with $N \geq 1$ are negative,
only  $F_0^{NS}(y\,  |     \,  \mu)$ survives in the 
asymptotic limit $\mu \to \infty$
while all the moments  $F_n^{NS}(y\,  |     \,  \mu)$
with $n \geq 1$ evolve to zero values.
Hence, in the formal $\mu \to \infty$
limit, we have  $$F^{NS}(x,y \,  |     
\, \mu \to \infty)\sim \delta(x)y \bar y$$ $i.e.,$ 
in each of its variables, the limiting function 
 $F^{NS}(x,y\,  |     \,  \mu \to \infty)$  
acquires the characteristic asymptotic form dictated by
the nature of the variable:
$\delta(x)$ is specific for the distribution functions \cite{gw,gp},
while  the $y \bar y$-form  is  
the asymptotic shape   for the lowest-twist two-body 
distribution amplitudes \cite{tmf,bl}.
For the nonforward distribution of a valence quark $q$ this gives
$${\cal F}^{val \, ;\,   q}_{\zeta}(X \, | \, \mu \to \infty) =6 N_q 
 X (1-X/\zeta) /\zeta^2 \, , $$
where $N_q$ is the  number of the valence $q$-quarks
in the hadron.

Another example  is  the evolution of the gluon distribution 
in pure gluodynamics which is governed by
$\gamma_{n+k}^{GG}$ with $\beta_0 = 11 N_c/3$.
Note that the lowest local operator in this case corresponds
to $n=1$. Furthermore, in pure gluodynamics,
$\gamma_{1}^{GG}$ vanishes  while $\gamma_{N}^{GG}<0$
if $N \geq 2$.
This means that in the $\mu \to \infty$ limit we have
 $$x F^{G}(x,y \,  |     
\, \mu \to \infty) =30 \, \delta(x) (y \bar y)^2$$ 
for the double distribution which results in 
$${\cal F}^{G}_{\zeta}(X \, | \,  \mu \to \infty) =30 \, 
 X^2 (1-X/\zeta)^2 /\zeta^3 $$
for the nonforward distribution.
In the formulas above, the total momentum carried 
by the gluons (in pure gluodynamics!) was normalized to unity.

In QCD,  we should take into account the effects
due to quark-gluon mixing.
Diagonalizing Eq.(\ref{a27}),  we obtain 
two multiplicatively renormalizable
combinations
 \begin{equation} 
F_{nk}^{\pm} = F_{nk}^Q + \alpha_{nk}^{\pm}  F_{nk}^G 
\end{equation}
where  (omitting the $nk$ indices) 
\begin{equation} 
 \alpha^{\pm} = \frac1{2 \gamma^{GQ}} \left (\gamma^{GG} - \gamma^{QQ} \pm
\sqrt{(\gamma^{GG} - \gamma^{QQ})^2 + 
4 \gamma^{GQ} \gamma^{QG}} \right ) \, .
\end{equation}
Their evolution is governed by the anomalous dimensions
\begin{equation} 
 \gamma^{\pm} = \frac12 \left ( \gamma^{GG} + \gamma^{QQ} \pm
\sqrt{(\gamma^{GG} - \gamma^{QQ})^2 + 
4 \gamma^{GQ} \gamma^{QG}} \right ) \, .
\end{equation}
In particular, $\gamma^+_{10} =0$ and $\alpha^+_{10} =1$
which means that $F_{10}^+ \equiv F_{10}^Q + F_{10}^G$ does not evolve:
the total momentum carried by the partons is conserved.
Another multiplicatively renormalizable combination 
involving  $F_{10}^Q$ and $ F_{10}^G$ is 
$$F_{10}^- =  F_{10}^Q - \frac{C_F}{4N_f} F_{10}^G\, .$$
It vanishes in the $\mu \to \infty$ limit, and we have
 \begin{equation}
F_{10}^Q (\mu \to \infty) \to \frac{N_f}{4 C_F + N_f}   \  \ ;  \  \ 
F_{10}^G (\mu \to \infty) \to \frac{4C_F}{4 C_F + N_f} 
 \, .
\end{equation}
Since  all  the combinations $F_{nk}^{\pm}$ with $n+k \geq 2$
vanish in the $\mu \to \infty$ limit, we obtain 
\begin{equation} 
 xF^G(x,y \, | \, \mu \to \infty) \to  
30 \, \frac{4C_F}{4 C_F + N_f} \, \delta(x)
(y \bar y)^2   \  \ ;  \  \ xF^Q(x,y \, | \, \mu \to \infty) \to  
30 \, \frac{N_f}{4 C_F + N_f} \, \delta(x)
(y \bar y)^2 \, , 
\end{equation}
or
\begin{equation} 
F^Q(x,y \, | \, \mu \to \infty) \to  
- 30 \, \frac{N_f}{4 C_F + N_f} \, \delta'(x)
(y \bar y)^2 \, .
 \end{equation}
In terms of nonforward distributions this is equivalent to
\begin{equation} 
 {\cal F}_{\zeta}^G (X \, | \, \mu \to \infty) \to 
30 \, \frac{4C_F}{4 C_F + N_f} \, \frac{X^2}{\zeta^3} 
\left (1- \frac{X}{\zeta} \right )^2 \, ,
\end{equation}
\begin{equation} 
 {\cal F}_{\zeta}^Q (X \, | \, \mu \to \infty) \to 
60 \, \frac{N_f}{4 C_F + N_f} \, \frac{X}{\zeta^2} 
\left (1- \frac{X}{\zeta} \right )
\left ( \frac{2 X}{\zeta} -1 \right )\, .
\end{equation}
Note that   both ${\cal F}^{Q}_{\zeta}(\zeta)$
and ${\cal F}^{G}_{\zeta}(\zeta)$ vanish in the $\mu \to \infty$
limit. 

\end{appendix}

\begin{figure}[htb]
\mbox{
   \epsfxsize=9cm
 \epsfysize=5cm
 \hspace{5cm}  
  \epsffile{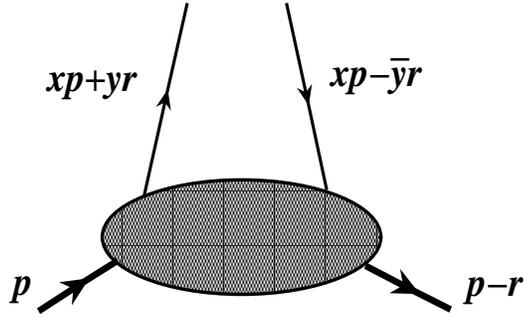}  }
  \vspace{0.5cm}
{\caption{\label{fig:double}   Parton picture 
for double distributions.
   }}
\end{figure}

\begin{figure}[htb]
\mbox{
   \epsfxsize=18cm
 \epsfysize=5cm  \epsffile{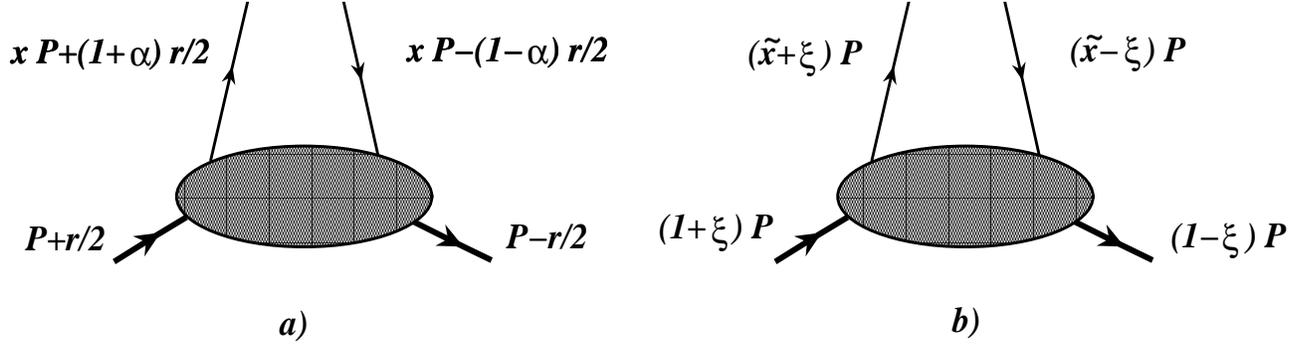}  }
{\caption{\label{fig:doubleji}   Parton picture in terms of 
 $a)$ modified double distributions and 
$b)$ off-forward parton distributions $H(\tilde x, \xi)$.
   }}
\end{figure}

\begin{figure}[htb]
\mbox{
   \epsfxsize=9cm
 \epsfysize=5cm
 \hspace{4.5cm}  
  \epsffile{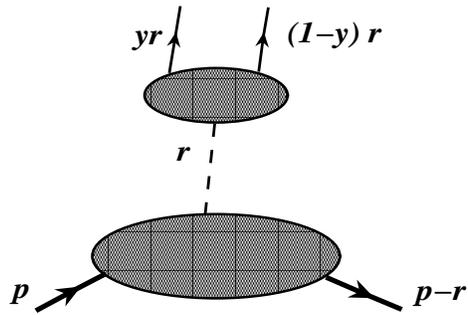}  }
  \vspace{0.5cm}
{\caption{\label{fig:doublemes}   Meson-like contribution.
   }}
\end{figure}

\end{document}